\documentclass[twocolumn]{aastex62}
\usepackage{amsmath}
\usepackage{graphicx}
\usepackage{natbib}
\usepackage[scaled]{helvet}
\usepackage{epsfig}
\usepackage{url}
\bibpunct{(}{)}{;}{a}{}{,}
\newcommand{\kms}{\,km\,s$^{-1}$}
\usepackage{textcomp}
\usepackage{mathpazo}
\usepackage{xspace}
\usepackage{hyperref}
\usepackage{apjfonts}
\usepackage{mathrsfs}
\usepackage{verbatim}

\usepackage{color}
\usepackage[normalem]{ulem}
\newcommand{\bjdtdb}{\ensuremath{\rm {BJD_{TDB}}}}
\newcommand{\feh}{\ensuremath{\left[{\rm Fe}/{\rm H}\right]}}

\newcommand{\teff}{\ensuremath{T_{\rm eff}}\xspace}

\newcommand{\msun}{\ensuremath{\,M_\Sun}}
\newcommand{\rsun}{\ensuremath{\,R_\Sun}}
\newcommand{\lsun}{\ensuremath{\,L_\Sun}}
\newcommand{\mj}{\ensuremath{\,M_{\rm J}}}
\newcommand{\rj}{\ensuremath{\,R_{\rm J}}}

\newcommand{\re}{\ensuremath{\,R_{\rm \Earth}}\xspace}

\newcommand{\fave}{\langle F \rangle}
\newcommand{\fluxcgs}{10$^9$ erg s$^{-1}$ cm$^{-2}$}

\newcommand{\loggstar}{\ensuremath{\log{g_\star}}}
\newcommand{\vsini}{\ensuremath{v\sin{I_*}}}
\newcommand{\ms}{\,m\,s$^{-1}$}
\newcommand{\thisstar}{KELT-24\xspace}
\newcommand{\thisplanet}{KELT-24\,b\xspace}

\newcommand{\rstar}{\ensuremath{R_{*}}}

\newcommand{\be}{\begin{equation}}
\newcommand{\ee}{\end{equation}}

\newcommand{\starmass}{$1.460^{+0.055}_{-0.059}$}
\newcommand{\starradius}{$1.506\pm0.022$}
\newcommand{\starage}{$0.78^{+0.61}_{-0.42}$}
\newcommand{\starvsini}{$19.76\pm0.160$}
\newcommand{\pmass}{$5.18^{+0.21}_{-0.22}$}
\newcommand{\prad}{$1.272\pm0.021$}
\newcommand{\pecc}{$0.077^{+0.024}_{-0.025}$}
\newcommand{\plambda}{$2.6^{+5.1}_{-3.6}$}

\begin{document}

\title{KELT-24b: A 5M$_{\rm J}$ Planet on a 5.6 day Well-Aligned Orbit around the Young V=8.3 F-star HD 93148}

\newcommand{\cfa}{Center for Astrophysics \textbar \ Harvard \& Smithsonian, 60 Garden St, Cambridge, MA 02138, USA}
\newcommand{\umich}{Astronomy Department, University of Michigan, 1085 S University Avenue, Ann Arbor, MI 48109, USA}
\newcommand{\utaustin}{Department of Astronomy, The University of Texas at Austin, Austin, TX 78712, USA}
\newcommand{\MIT}{Department of Physics and Kavli Institute for Astrophysics and Space Research, Massachusetts Institute of Technology, Cambridge, MA 02139, USA}
\newcommand{\MITEPS}{Department of Earth, Atmospheric and Planetary Sciences, Massachusetts Institute of Technology,  Cambridge,  MA 02139, USA}
\newcommand{\uflorida}{Department of Astronomy, University of Florida, 211 Bryant Space Science Center, Gainesville, FL, 32611, USA}
\newcommand{\riverside}{Department of Earth Sciences, University of California,
Riverside, CA 92521, USA}
\newcommand{\usq}{University of Southern Queensland, Centre for Astrophysics, West Street, Toowoomba, QLD 4350 Australia}
\newcommand{\ames}{NASA Ames Research Center, Moffett Field, CA, 94035, USA}
\newcommand{\geneva}{Observatoire de l’Universit\'e de Gen\`eve, 51 chemin des Maillettes,
1290 Versoix, Switzerland}
\newcommand{\uw}{Astronomy Department, University of Washington, Seattle, WA 98195 USA}
\newcommand{\warwick}{Department of Physics, University of Warwick, Gibbet Hill Road, Coventry CV4 7AL, UK}
\newcommand{\warwickceh}{Centre for Exoplanets and Habitability, University of Warwick, Gibbet Hill Road, Coventry CV4 7AL, UK}
\newcommand{\princeton}{Department of Astrophysical Sciences, Princeton University, 4 Ivy Lane, Princeton, NJ, 08544, USA}
\newcommand{\liege}{Space Sciences, Technologies and Astrophysics Research (STAR) Institute, Universit\'e de Li\`ege, 19C All\'ee du 6 Ao\^ut, 4000 Li\`ege, Belgium}
\newcommand{\vanderbilt}{Department of Physics and Astronomy, Vanderbilt University, Nashville, TN 37235, USA}
\newcommand{\fisk}{Department of Physics, Fisk University, 1000 17th Avenue North, Nashville, TN 37208, USA}
\newcommand{\columbia}{Department of Astronomy, Columbia University, 550 West 120th Street, New York, NY 10027, USA}
\newcommand{\toronto}{Dunlap Institute for Astronomy and Astrophysics, University of Toronto, Ontario M5S 3H4, Canada}
\newcommand{\unc}{Department of Physics and Astronomy, University of North Carolina at Chapel Hill, Chapel Hill, NC 27599, USA}
\newcommand{\iac}{Instituto de Astrof\'isica de Canarias (IAC), E-38205 La Laguna, Tenerife, Spain}
\newcommand{\lalaguna}{Departamento de Astrof\'isica, Universidad de La Laguna (ULL), E-38206 La Laguna, Tenerife, Spain}
\newcommand{\louisville}{Department of Physics and Astronomy, University of Louisville, Louisville, KY 40292, USA}
\newcommand{\aavso}{American Association of Variable Star Observers, 49 Bay State Road, Cambridge, MA 02138, USA}
\newcommand{\utokyo}{The University of Tokyo, 7-3-1 Hongo, Bunky\={o}, Tokyo 113-8654, Japan}
\newcommand{\naoj}{National Astronomical Observatory of Japan, 2-21-1 Osawa, Mitaka, Tokyo 181-8588, Japan}
\newcommand{\jstpresto}{JST, PRESTO, 2-21-1 Osawa, Mitaka, Tokyo 181-8588, Japan}
\newcommand{\astrobiojapan}{Astrobiology Center, 2-21-1 Osawa, Mitaka, Tokyo 181-8588, Japan}
\newcommand{\ctio}{Cerro Tololo Inter-American Observatory, Casilla 603, La Serena, Chile}
\newcommand{\nexsci}{Caltech IPAC -- NASA Exoplanet Science Institute 1200 E. California Ave, Pasadena, CA 91125, USA}
\newcommand{\ucsc}{Department of Astronomy and Astrophysics, University of
California, Santa Cruz, CA 95064, USA}
\newcommand{\gsfc}{Exoplanets and Stellar Astrophysics Laboratory, Code 667, NASA Goddard Space Flight Center, Greenbelt, MD 20771, USA}
\newcommand{\sgtinc}{SGT, Inc./NASA AMES Research Center, Mailstop 269-3, Bldg T35C, P.O. Box 1, Moffett Field, CA 94035, USA}
\newcommand{\chile}{Center of Astro-Engineering UC, Pontificia Universidad Cat\'olica de Chile, Av. Vicu\~{n}a Mackenna 4860, 7820436 Macul, Santiago, Chile}
\newcommand{\Pontificia}{Instituto de Astrof\'isica, Pontificia Universidad Cat\'olica de Chile, Av.\ Vicu\~na Mackenna 4860, Macul, Santiago, Chile}
\newcommand{\Millennium}{Millennium Institute for Astrophysics, Chile}
\newcommand{\maxplank}{Max-Planck-Institut f\"ur Astronomie, K\"onigstuhl 17, Heidelberg 69117, Germany}
\newcommand{\utdallas}{Department of Physics, The University of Texas at Dallas, 800 West
Campbell Road, Richardson, TX 75080-3021 USA}
\newcommand{\MauryLewin}{Maury Lewin Astronomical Observatory, Glendora, CA 91741, USA}
\newcommand{\umbc}{University of Maryland, Baltimore County, 1000 Hilltop Circle, Baltimore, MD 21250, USA}
\newcommand{\osu}{Department of Astronomy, The Ohio State University, 140 West 18th Avenue, Columbus, OH 43210, USA}
\newcommand{\MITAA}{Department of Aeronautics and Astronautics, MIT, 77 Massachusetts Avenue, Cambridge, MA 02139, USA}
\newcommand{\openu}{School of Physical Sciences, The Open University, Milton Keynes MK7 6AA, UK}
\newcommand{\swarthmore}{Department of Physics and Astronomy, Swarthmore College, Swarthmore, PA 19081, USA}
\newcommand{\seti}{SETI Institute, Mountain View, CA 94043, USA}
\newcommand{\lehigh}{Department of Physics, Lehigh University, 16 Memorial Drive East, Bethlehem, PA 18015, USA}
\newcommand{\utah}{Department of Physics and Astronomy, University of Utah, 115 South 1400 East, Salt Lake City, UT 84112, USA}
\newcommand{\USNA}{Department of Physics, United States Naval Academy, 572C Holloway Rd., Annapolis, MD 21402, USA}
\newcommand{\DTM}{Department of Terrestrial Magnetism, Carnegie Institution for Science, 5241 Broad Branch Road, NW, Washington, DC 20015, USA}
\newcommand{\UPenn}{The University of Pennsylvania, Department of Physics and Astronomy, Philadelphia, PA, 19104, USA}
\newcommand{\montana}{Department of Physics and Astronomy, University of Montana, 32 Campus Drive, No. 1080, Missoula, MT 59812 USA}
\newcommand{\psu}{Department of Astronomy \& Astrophysics, The Pennsylvania State University, 525 Davey Lab, University Park, PA 16802, USA}
\newcommand{\psust}{Center for Exoplanets and Habitable Worlds, The Pennsylvania State University, 525 Davey Lab, University Park, PA 16802, USA}
\newcommand{\Kutztown}{Department of Physical Sciences, Kutztown University, Kutztown, PA 19530, USA}
\newcommand{\udel}{Department of Physics \& Astronomy, University of Delaware, Newark, DE 19716, USA}
\newcommand{\Westminster}{Department of Physics, Westminster College, New Wilmington, PA 16172}
\newcommand{\steward}{Department of Astronomy and Steward Observatory, University of Arizona, Tucson, AZ 85721, USA}
\newcommand{\saao}{South African Astronomical Observatory, PO Box 9, Observatory, 7935, Cape Town, South Africa}
\newcommand{\salt}{Southern African Large Telescope, PO Box 9, Observatory, 7935, Cape Town, South Africa}
\newcommand{\ssl}{Societ\`{a} Astronomica Lunae, Italy}
\newcommand{\spot}{Spot Observatory, Nashville, TN 37206, USA}
\newcommand{\txamGP}{George P.\ and Cynthia Woods Mitchell Institute for Fundamental Physics and Astronomy, Texas A\&M University, College Station, TX77843 USA}
\newcommand{\txam}{Department of Physics and Astronomy, Texas A\&M university, College Station, TX 77843 USA}
\newcommand{\wellesley}{Department of Astronomy, Wellesley College, Wellesley, MA 02481, USA}
\newcommand{\byu}{Department of Physics and Astronomy, Brigham Young University, Provo, UT 84602, USA}
\newcommand{\Hazelwood}{Hazelwood Observatory, Churchill, Victoria, Australia}
\newcommand{\pest}{Perth Exoplanet Survey Telescope}
\newcommand{\Winer}{Winer Observatory, PO Box 797, Sonoita, AZ 85637, USA}
\newcommand{\icpo}{Ivan Curtis Private Observatory}
\newcommand{\elsauce}{El Sauce Observatory, Chile}
\newcommand{\crow}{Atalaia Group \& CROW Observatory, Portalegre, Portugal}
\newcommand{\dfus}{Dipartimento di Fisica ``E.R.Caianiello'', Universit\`a di Salerno, Via Giovanni Paolo II 132, Fisciano 84084, Italy}
\newcommand{\indfn}{Istituto Nazionale di Fisica Nucleare, Napoli, Italy}
\newcommand{\sotes}{Gabriel Murawski Private Observatory (SOTES)}
\newcommand{\stromlo}{Research School of Astronomy and Astrophysics, Mount Stromlo Observatory, Australian National University, Cotter Road, Weston, ACT, 2611, Australia}
\newcommand{\wyoming}{Department of Physics \& Astronomy, University of Wyoming, 1000 E University Ave, Dept 3905, Laramie, WY 82071, USA}
\newcommand{\thai}{National Astronomical Research Institute of Thailand, 260, Moo 4, T. Donkaew, A. Mae Rim, Chiang Mai, 50180, Thailand}
\newcommand{\asp}{Acton Sky Portal (private observatory), Acton, MA 01720,  USA}
\newcommand{\kyoto}{Department of Physics, Faculty of Science, Kyoto Sangyo University, Kamigamo Motoyama, Kita-ku, Kyoto, 603-8555, Japan}
\newcommand{\chiba}{Planetary Exploration Research Center, Chiba Institute of Technology, 2-17-1 Tsudanuma, Narashino, Chiba 275-0016, Japan}
\newcommand{\berkely}{Department of Astronomy, University of California Berkeley, Berkeley, CA 94720-3411, USA}
\newcommand{\hawaii}{Institute for Astronomy, University of Hawai'i, 2680 Woodlawn Drive, Honolulu, HI 96822, USA}
\newcommand{\carnegie}{The Observatories of the Carnegie Institution for Science, 813 Santa Barbara St., Pasadena, CA 91101, USA}
\newcommand{\brazil}{Instituto de Astronomia, Geof\'isica e Ciencias Atmosf\'ericas, Universidade de S\`ao Paulo, Rua do Mat\~ao 1226, Cidade Universit\~aria, S\'ao Paulo, SP 05508-900, Brazil}
\newcommand{\bhicfa}{Black Hole Initiative at Harvard University, 20 Garden Street, Cambridge, MA 02138, USA}
\newcommand{\FFL}{\altaffiliation{Future Faculty Leaders Fellow}}
\newcommand{\torres}{\altaffiliation{Juan Carlos Torres Fellow}}
\newcommand{\sagan}{\altaffiliation{NASA Sagan Fellow}}
\newcommand{\bernoulli}{\altaffiliation{Bernoulli fellow}}
\newcommand{\gruber}{\altaffiliation{Gruber fellow}}
\newcommand{\kavli}{\altaffiliation{Kavli Fellow}}
\newcommand{\peg}{\altaffiliation{51 Pegasi b Fellow}}
\newcommand{\pappalardo}{\altaffiliation{Pappalardo Fellow}}
\newcommand{\hubble}{\altaffiliation{NASA Hubble Fellow}}
\newcommand{\carnegiefellow}{\altaffiliation{Carnegie Fellow}}

\correspondingauthor{Joseph E. Rodriguez} 
\email{joseph.rodriguez@cfa.harvard.edu}

\author[0000-0001-8812-0565]{Joseph E. Rodriguez} 
\FFL
\affiliation{\cfa}

\author[0000-0003-3773-5142]{Jason D. Eastman} 
\affiliation{\cfa}

\author[0000-0002-4891-3517]{George Zhou} 
\hubble
\affiliation{\cfa}

\author[0000-0002-8964-8377]{Samuel N. Quinn} 
\affiliation{\cfa}

\author[0000-0002-9539-4203]{Thomas G.\ Beatty}
\affiliation{\steward}

\author[0000-0003-4464-1371]{Kaloyan Penev} 
\affiliation{\utdallas}

\author[0000-0002-5099-8185]{Marshall C.\ Johnson}
\affiliation{\osu}

\author[0000-0002-1617-8917]{Phillip A. Cargile}
\affiliation{\cfa}

\author[0000-0001-9911-7388]{David W. Latham} 
\affiliation{\cfa}

\author[0000-0001-6637-5401]{Allyson Bieryla} 
\affiliation{\cfa}

\author[0000-0001-6588-9574]{Karen A. Collins} 
\affiliation{\cfa}

\author[0000-0001-8189-0233]{Courtney D. Dressing}
\affiliation{\berkely}

\author[0000-0002-5741-3047]{David R. Ciardi}  
\affiliation{\nexsci}

\author{Howard M. Relles} 
\affiliation{\cfa}

\author[0000-0001-7809-1457]{Gabriel Murawski}
\affiliation{\sotes}

\author[0000-0003-1510-8981]{Taku Nishiumi}
\affiliation{\kyoto}
\affiliation{\naoj}

\author[0000-0003-3480-0973]{Atsunori Yonehara}
\affiliation{\kyoto}

\author[0000-0001-7918-118X]{Ryo Ishimaru}
\affiliation{\chiba}

\author[0000-0002-3286-911X]{Fumi Yoshida}
\affiliation{\chiba}

\author{Joao Gregorio}
\affiliation{\crow}

\author[0000-0003-2527-1598]{Michael B. Lund}
\affiliation{\nexsci}

\author[0000-0002-5951-8328]{Daniel J.\ Stevens}
\affiliation{\psu}
\affiliation{\psust}

\author[0000-0002-3481-9052]{Keivan G. Stassun} 
\affiliation{\vanderbilt}
\affiliation{\fisk}

\author[0000-0003-0395-9869]{B. Scott Gaudi} 
\affiliation{\osu}

\author[0000-0001-8020-7121]{Knicole D.\ Col\'on}
\affiliation{\gsfc}

\author[0000-0002-3827-8417]{Joshua Pepper} 
\affiliation{\lehigh}

\author[0000-0001-8511-2981]{Norio Narita}
\affiliation{\astrobiojapan}
\affiliation{\jstpresto}
\affiliation{\naoj}
\affiliation{\iac}

\author[0000-0003-3251-3583]{Supachai Awiphan}
\affiliation{\thai}

\author[0000-0001-6327-1113]{Pongpichit Chuanraksasat}
\affiliation{\thai}

\author{Paul\ Benni}
\affiliation{\asp}

\author{Roberto Zambelli}
\affiliation{\ssl}

\author[0000-0002-9853-5673]{Lehman H. Garrison}
\affiliation{\cfa}

\author[0000-0003-1928-0578]{Maurice L. Wilson} 
\affiliation{\cfa}

\author[0000-0003-1012-4771]{Matthew A. Cornachione}
\affiliation{\utah}
\affiliation{\USNA}

\author[0000-0002-6937-9034]{Sharon X. Wang}
\affiliation{\DTM}

\author[0000-0002-2919-6786]{Jonathan Labadie-Bartz}
\affiliation{\brazil}

\author{Romy Rodr\'iguez}
\affiliation{\osu}

\author[0000-0001-5016-3359]{Robert J.\ Siverd}
\affiliation{\vanderbilt}

\author[0000-0003-4554-5592]{Xinyu Yao}
\affiliation{\lehigh}


\author{Daniel Bayliss}
\affiliation{\warwick}
\affiliation{\warwickceh}

\author{Perry Berlind} 
\affiliation{\cfa}


\author[0000-0002-2830-5661]{Michael L. Calkins} 
\affiliation{\cfa}

\author[0000-0002-8035-4778]{Jessie L. Christiansen}
\affiliation{\nexsci}

\author[0000-0003-2995-4767]{David H.\ Cohen}
\affiliation{\swarthmore}

\author[0000-0003-2239-0567]{Dennis M.\ Conti} 
\affiliation{\aavso}

\author{Ivan A.\ Curtis}
\affiliation{\icpo}

\author{D.\ L.\ Depoy}
\affiliation{\txamGP}
\affiliation{\txam}

\author{Gilbert A. Esquerdo} 
\affiliation{\cfa}

\author[0000-0002-5674-2404]{Phil Evans}
\affiliation{\elsauce}

\author[0000-0002-2457-7889]{Dax Feliz}
\affiliation{\vanderbilt}

\author{Benjamin J.\ Fulton}
\affiliation{\nexsci}

\author[0000-0001-9206-3460]{Thomas~W.-S.~Holoien}
\carnegiefellow
\affiliation{\carnegie}

\author[0000-0001-5160-4486]{David J. James}
\affiliation{\cfa}
\affiliation{\bhicfa}

\author[0000-0002-6244-477X]{Tharindu Jayasinghe}
\affiliation{\osu}

\author{Hannah Jang-Condell}
\affiliation{\wyoming}

\author[0000-0002-4625-7333]{Eric L.\ N.\ Jensen}
\affiliation{\swarthmore}

\author[0000-0002-1704-6289]{John A. Johnson}
\affiliation{\cfa}

\author[0000-0003-0634-8449]{Michael D.\ Joner}
\affiliation{\byu}

\author[0000-0002-1910-7065]{Somayeh Khakpash}
\affiliation{\lehigh}

\author[0000-0003-0497-2651]{John F.\ Kielkopf} 
\affiliation{\louisville}

\author[0000-0002-4236-9020]{Rudolf B. Kuhn}
\affiliation{\saao}
\affiliation{\salt}

\author{Mark Manner}
\affiliation{\spot}

\author{Jennifer L.\ Marshall}  
\affiliation{\txamGP}
\affiliation{\txam}

\author[0000-0001-9504-1486]{Kim K.\ McLeod}
\affiliation{\wellesley}

\author[0000-0002-8041-1832]{Nate McCrady}
\affiliation{\montana}

\author{Thomas E.\ Oberst}
\affiliation{\Westminster}

\author[0000-0002-0582-1751]{Ryan J. Oelkers} 
\affiliation{\vanderbilt}

\author[0000-0001-7506-5640]{Matthew T.\ Penny}
\affiliation{\osu}

\author[0000-0002-5005-1215]{Phillip A.\ Reed}
\affiliation{\Kutztown}

\author{David H. Sliski}
\affiliation{\UPenn}

\author[0000-0003-4631-1149]{B.~J.~Shappee}
\affiliation{\hawaii}

\author{Denise C.\ Stephens}
\affiliation{\byu}

\author[0000-0003-2163-1437]{Chris Stockdale}
\affiliation{\Hazelwood}

\author[0000-0001-5603-6895]{Thiam-Guan Tan}
\affiliation{\pest}

\author[0000-0002-5867-082X]{Mark Trueblood}  
\affiliation{\Winer}

\author{Pat Trueblood}  
\affiliation{\Winer}

\author[0000-0001-6213-8804]{Steven~Villanueva~Jr.} 
\pappalardo
\affiliation{\MIT}

\author[0000-0001-9957-9304]{Robert A. Wittenmyer}
\affiliation{\usq}

\author[0000-0001-6160-5888]{Jason T. Wright}
\affiliation{\psu}
\affiliation{\psust}

\shorttitle{Go KELT! Go KELT-24\MakeLowercase{b}!}
\shortauthors{Rodriguez et al.}

\begin{abstract}

We present the discovery of KELT-24 b, a massive hot Jupiter orbiting a bright (V=8.3 mag, K=7.2 mag) young F-star with a period of 5.6 days. The host star, KELT-24 (HD 93148), has a $T_{\rm eff}$ =$6509^{+50}_{-49}$~K, a mass of $M_{*}$ = $1.460^{+0.055}_{-0.059}$ $M_{\odot}$, radius of $R_{*}$ = $1.506\pm0.022$ $R_{\odot}$, and an age of $0.78^{+0.61}_{-0.42}$ Gyr. Its planetary companion (KELT-24 b) has a radius of $R_{\rm P}$ = $1.272\pm0.021$ $R_{\rm J}$, a mass of $M_{\rm P}$ = $5.18^{+0.21}_{-0.22}$ $M_{\rm J}$, and from Doppler tomographic observations, we find that the planet's orbit is well-aligned to its host star's projected spin axis ($\lambda$ = $2.6^{+5.1}_{-3.6}$). The young age estimated for KELT-24 suggests that it only recently started to evolve from the zero-age main sequence. KELT-24 is the brightest star known to host a transiting giant planet with a period between 5 and 10 days. Although the circularization timescale is much longer than the age of the system, we do not detect a large eccentricity or significant misalignment that is expected from dynamical migration. The brightness of its host star and its moderate surface gravity make KELT-24b an intriguing target for detailed atmospheric characterization through spectroscopic emission measurements since it would bridge the current literature results that have primarily focused on lower mass hot Jupiters and a few brown dwarfs.

\end{abstract}


\section{Introduction} 
Despite confirmation of 4000 planets orbiting other stars, many of the questions raised by the first few discoveries over 20 years ago remain unanswered. One of the first possible planetary systems ever discovered was HD 114762 b, a massive Jupiter on an 84 day period around a late F-star \citep{Latham:1989}. The inclination of the companion's orbit is not known, but it has a minimum mass of 11 \mj\ \citep{Latham:1989}. Interestingly, over the past 30 years since this discovery, we now know of over 250 planets with a measured mininum mass between 4 and 13 Jupiter masses. Above $\sim$13\mj, a sub-stellar companion can begin to fuse deuterium in its core, currently an arbitrary method for distinguishing planets and brown dwarfs. Another method to distinguish between brown dwarfs and giant planets is their formation mechanisms. Formation theories for brown dwarfs are similar to stars, in that they form either through gravitational instability or molecular cloud fragmentation while gas giant planet formation is likely dominated by core accretion \citep[and references therein]{Chabrier:2014}. However, there are inconsistencies between the deuterium burning and formation arguments to distinguish between planets and brown dwarfs since it is possible to form an object above the deuterium-burning limit through core accretion \citep{Baraffe:2008,Molliere:2012,Bodenheimer:2013}. The distinction between brown dwarfs and giant planets has been debated for decades \citep[e.g.][]{Chabrier:2003, Chabrier:2007, Spiegel:2011}. The detailed characterization of massive Jupiters and low-mass brown dwarfs may shed light on their defining characteristics. However, the relatively low number of massive giant planets transiting bright host stars known, combined with their expected smaller atmospheric scale heights due to their higher surface gravity, has so far limited atmospheric studies of massive hot Jupiters and brown dwarfs.

The discovery of 51 Peg b \citep{Mayor:1995}, a Jupiter-mass object orbiting a Sun-like star with a period of only 4.23 days, led to the idea that giant planets must undergo large scale migration since it was commonly believed that giant planets could only form out past the ice line. However, it is not clear what mechanisms govern this migration, or if giant planets can form in situ close to the star \citep{Batygin:2016, Huang:2016}. One possibility is that giant planets migrate slowly and smoothly within the circumstellar gas-dust disk, resulting in well-aligned, nearly circular orbits \citep{Goldreich:1980, Lin:1996, Dangelo:2003}. It is also thought that planetary migration may be heavily influenced by gravitational interactions with other bodies within the system.  These interactions can result in highly eccentric and misaligned orbits (relative to the rotation axis of the star), that will dampen due to tidal effects, and is typically referred to as ``High Eccentricity Migration (HEM)" \citep{Rasio:1996, Wu:2003, Fabrycky:2007, Nagasawa:2011, Wu:2011}. 


Due to their high rotation velocities (\vsini > 10 \kms), hot ($>$ 6250 K) massive stars were avoided by many spectroscopic and photometric surveys for planets, including {\it Kepler} \citep{Borucki:2010}. This was primarily due to the difficulty in measuring precise radial velocities ($<$200\ms) from rotationally broadened spectral lines. However, with the advancement of techniques444 to measure the Rossiter-McLaughlin effect \citep{Rossiter:1924, McLaughlin:1924, Gaudi:2007, Cegla:2016} and Doppler Tomography \citep{CollierCameron:2010}, a few dozen giant planets have now been confirmed around rapidly rotating F- and A-type stars. From these discoveries, a pattern has emerged where hot Jupiters around massive stars tend to be in misaligned orbits relative to their host star's rotation axis \citep{Winn:2010b, Albrecht:2012}. This observed trend might be a signature that hot Jupiters predominantly migrate through HEM resulting in highly-eccentric and misaligned short-period orbits (see review \citealp{Dawson:2018} for a more in-depth discussion on tidal migration of hot Jupiters.).

\begin{figure}[ht!]
    \vspace{.0in}
    \centering\includegraphics[width=0.99\columnwidth, trim = 0.5in 2.9in 0.5in 0.3in]{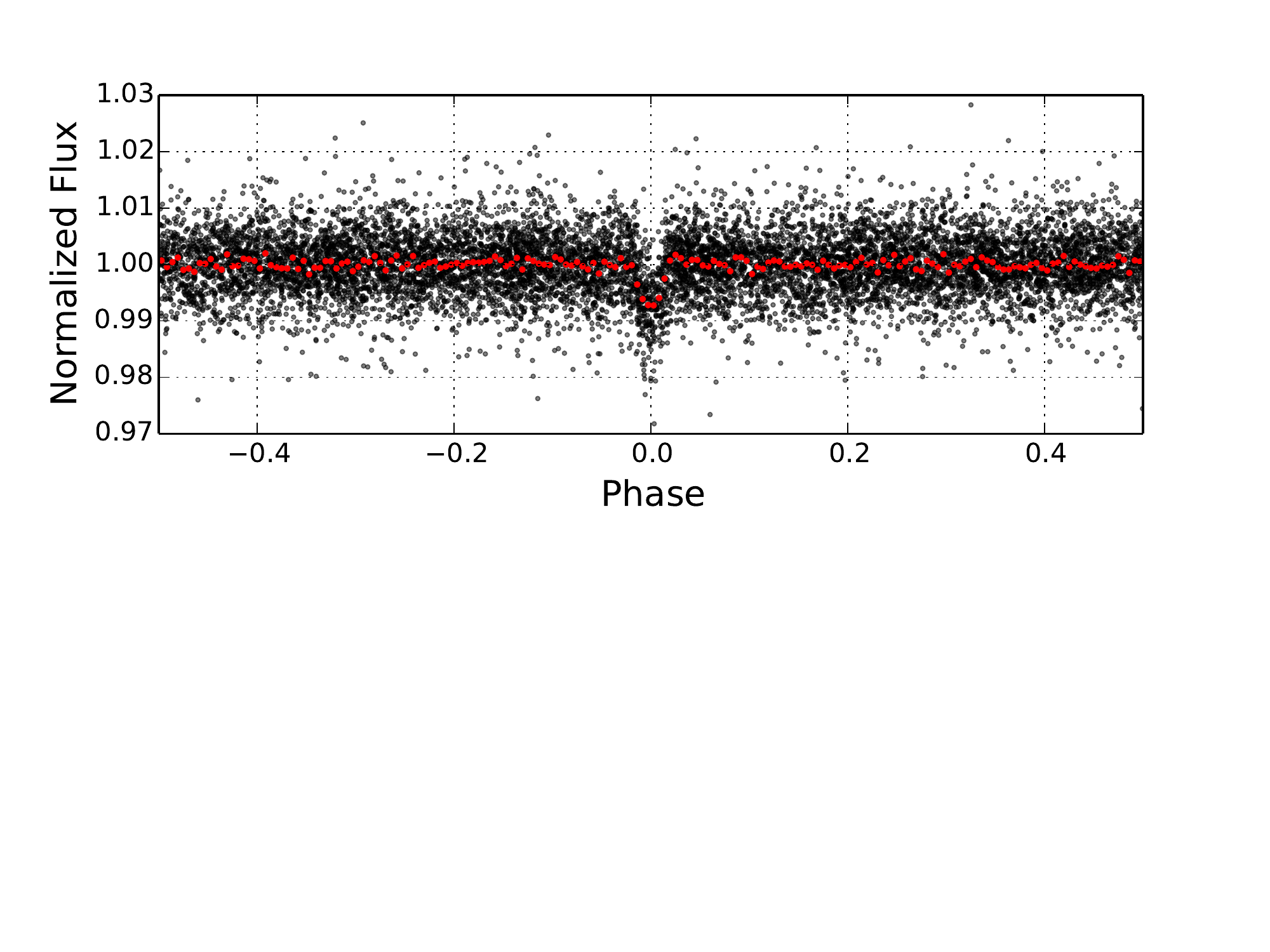}
    \caption{The KELT discovery light curve of \thisstar\ containing 10181 observations from the KELT-North telescope phase-folded on the discovery period of 5.551477 days. The red points are the data binned on a 45-minute time scale. }
    \label{fig:DiscoveryLC}
\end{figure}

With the recent launch and early success of NASA's Transiting Exoplanet Survey Satellite ({\it TESS}) mission, we have now entered the next major chapter in the field of exoplanets. The primary mission for {\it TESS} is to discover and measure the masses of 50 small planets (R$ < 4 \re$) to understand their bulk compositions \citep{Ricker:2015}. However, {\it TESS} will observe over 85\% of the entire sky during its 2-year nominal mission, with the expectation of discovering thousands of giant planets \citep{Barclay:2018}. There have already been over a dozen discoveries announced, including a number of giant planets on short-period, eccentric orbits \citep{Brahm:2018, Nielsen:2019, Rodriguez:2019}. Also, unlike the ground-based surveys which struggle to discover longer period hot Jupiters due to their poor duty cycle, {\it TESS} should be complete for all short-period transiting hot Jupiters (P $\lesssim$ 5\ days). Short-duration ground-based surveys can be a great asset in the discovery of longer period giant planets (P $\gtrsim$ 5\ days) by precovering the ephemerides of the hundreds of single transits expected in the short baseline ($\sim$27 days) of {\it TESS} observations \citep{Villanueva:2019, Yao:2019}. Additionally, detecting planets before their host stars are observed by {\it TESS} provides the opportunity to take advantage of the photometric precision and conduct detailed characterization of the planet's atmosphere through optical phase curves (see WASP-18b, \citealp{Shporer:2019}).

In this paper, we present the discovery of KELT-24b, a massive (\pmass\ \mj) Jupiter on a prograde orbit ($\lambda$ = \plambda\ degrees) orbiting a young (\starage\ Gyr) F-star. The brightness of the host star (V=8.3 mag) and the large planetary radius (1.2 R$_J$) relative to its high mass, makes \thisplanet well-suited for detailed characterization of its atmosphere. Additionally, \thisplanet should be observed by {\it TESS} in the upcoming sectors 20 and 21, which should provide a great opportunity for simultaneous observations with the Hubble Space Telescope. 


The paper is organized in the following way: We describe our observations and the detection of KELT-24b as a candidate in \S2. We present our global analysis of all observations in \S3. In \S4 we place the KELT-24 system in context with all known planets and given an overview of future detailed characterization observations for which it would be well-suited. We summarize our results and conclusions in \S5.


\begin{table}
\scriptsize
\setlength{\tabcolsep}{2pt}
\centering
\caption{Literature and Measured Properties for \thisstar}
\begin{tabular}{llcc}
  \hline
  \hline
Other identifiers\dotfill & \\
\multicolumn{3}{c}{HD 93148} \\
\multicolumn{3}{c}{HIP 52796,   TYC 4388-1652-1} \\
\multicolumn{3}{c}{BD+72 502,   TIC 349827430}\\
\hline
\hline
Parameter & Description & Value & Source\\
\hline 
$\alpha_{J2000}$\dotfill	&Right Ascension (RA)\dotfill & 10:47:38.35101& 1	\\
$\delta_{J2000}$\dotfill	&Declination (Dec)\dotfill & +71:39:21.15672& 1	\\
\\
$l$\dotfill     & Galactic Longitude\dotfill & 135.5728726$^\circ$ & 1\\
$b$\dotfill     & Galactic Latitude\dotfill & +42.30147339$^\circ$ & 1\\
\\
B$_T$\dotfill			&Tycho B$_T$ mag.\dotfill & 8.913$^{+0.020}_{-0.016}$		& 2	\\
V$_T$\dotfill			&Tycho V$_T$ mag.\dotfill & 8.389$^{+0.020}_{-0.012}$		& 2	\\
${\rm G}$\dotfill     & Gaia $G$ mag.\dotfill     &8.238$\pm$0.02& 1\\
\\
J\dotfill			& 2MASS J mag.\dotfill & 7.408 $\pm$ 0.020	& 3	\\
H\dotfill			& 2MASS H mag.\dotfill & 7.200 $\pm$ 0.04	    &  3	\\
K$_S$\dotfill			& 2MASS ${\rm K_S}$ mag.\dotfill & 7.154 $\pm$ 0.02&  3	\\
\\
\textit{WISE1}\dotfill		& \textit{WISE1} mag.\dotfill & 7.106 $\pm$ 0.039 & 4	\\
\textit{WISE2}\dotfill		& \textit{WISE2} mag.\dotfill & 7.134$^{+0.030}_{-0.019}$ &  4 \\
\textit{WISE3}\dotfill		& \textit{WISE3} mag.\dotfill &  7.148$^{+0.030}_{-0.017}$& 4	\\
\textit{WISE4}\dotfill		& \textit{WISE4} mag.\dotfill & 7.184$^{+0.1}_{-0.098}$ &  4	\\
\\
$\mu_{\alpha}$\dotfill		& Gaia DR2 proper motion\dotfill & -56.184 $\pm$ 0.053 & 1 \\
                    & \hspace{3pt} in RA (mas yr$^{-1}$)	&  \\
$\mu_{\delta}$\dotfill		& Gaia DR2 proper motion\dotfill 	&  -34.808 $\pm$ 0.064 &  1 \\
                    & \hspace{3pt} in DEC (mas yr$^{-1}$) &  \\
$\pi^\ddagger$\dotfill & Gaia Parallax (mas) \dotfill & 10.414 $\pm$  0.0469$^{\dagger}$ &  1 \\
$RV$\dotfill & Systemic radial \hspace{9pt}\dotfill  & $-5.749\pm0.065$  & \S\ref{sec:TRES} \\
     & \hspace{3pt} velocity (\kms)  & \\
$d$\dotfill & Distance (pc)\dotfill & 96.025$\pm0.306^{\ddagger}$ & 1 \\
$U^{*}$\dotfill & Space Velocity (\kms)\dotfill & $-11.00 \pm 0.11$  & \S\ref{sec:uvw} \\
$V$\dotfill       & Space Velocity (\kms)\dotfill & $-9.36 \pm 0.10$ & \S\ref{sec:uvw} \\
$W$\dotfill       & Space Velocity (\kms)\dotfill & $0.11\pm 0.05$ & \S\ref{sec:uvw} \\
\hline
\end{tabular}
\begin{flushleft}
 \footnotesize{ \textbf{\textsc{NOTES:}}
 The uncertainties of the photometry have a systematic error floor applied. \\
 $\dagger$ RA and Dec are in epoch J2000. The coordinates come from Vizier where the Gaia RA and Dec have been precessed to J2000 from epoch J2015.5.\\
 $\ddagger$ Values have been corrected for the -0.82 $\mu$as offset as reported by \citet{Stassun:2018}.\\
 $*$ $U$ is in the direction of the Galactic center. \\
 References are: $^1$\citet{Gaia:2018},$^2$\citet{Hog:2000},$^3$\citet{Cutri:2003}, $^4$\citet{Zacharias:2017}
}
\end{flushleft}
\label{tbl:LitProps}
\end{table}

\begin{table*}
 \centering
 \caption{Photometric follow-up observations of \thisplanet and the detrending parameters used for the global fit.}
 \label{tbl:detrending_parameters}
 \begin{tabular}{lllllllll}
    \hline
    \hline
Observatory & Date (UT) & Diameter (m) & Filter & FOV & Pixel Scale  & Exposure (s) & Detrending\\
    \hline
ULMT & 2019 March 30 & 0.6096 &$z^{\prime}$ & 26.8$\arcmin$ $\times$ 26.8$\arcmin$&  0.39$\arcsec$ & 100 & airmass, \textit{x} coordinates \\
FLWO/KeplerCam &2019 March 30 & 1.2 &$i^{\prime}$ & 23.1$\arcmin$ $\times$ 23.1$\arcmin$&  0.37$\arcsec$ & 60 & airmass \\
ULMT & 2019 April 11 & 0.6096 &$z^{\prime}$ & 26.8$\arcmin$ $\times$ 26.8$\arcmin$&  0.39$\arcsec$ & 100 &  None \\
FLWO/KeplerCam &2019 April 11 & 1.2 &$i^{\prime}$ & 23.1$\arcmin$ $\times$ 23.1$\arcmin$&  0.37$\arcsec$ & 90 & airmass \\
SOTES & 2019 April 16 & 0.08 & $R$ & 84$\arcmin$ $\times$ 57$\arcmin$ &   1.52$\arcsec$ & 240 & airmass \\
CROW & 2019 April 27 & 0.354 & $i^{\prime}$ & 23$\arcmin$ $\times$ 18$\arcmin$ &   0.66$\arcsec$ & 60 & airmass \\
LCO TFN & 2019 April 27 & 0.4 & $z^{\prime}$ & 19$\arcmin$ $\times$ 29$\arcmin$ &   0.57$\arcsec$ & 30 & airmass \\
KAO & 2019 May 03 & 1.3 & $z^{\prime}$ & 12.2$\arcmin$ $\times$ 12.2$\arcmin$ & 0.357$\arcsec$  & 40 & airmass\\
KCP & 2019 May 03 & 1.0 & $z^{\prime}$ & 12.2$\arcmin$ $\times$ 12.2$\arcmin$ & 0.24$\arcsec$  & 30 & airmass\\
        \hline
        \hline
 \end{tabular}
\begin{flushleft}
  \footnotesize \textbf{\textsc{NOTES:}} All the follow-up photometry presented in this paper is available in machine-readable form in the online journal. See \citet{Collins:2018} for detailed description of the KELT-FUN facilities.
  \end{flushleft}
\end{table*}

\section{Observations and Archival Data}
\label{Obs}

\subsection{KELT Photometry}
\label{sec:KELT}
The Kilodegree Extremely Little Telescope (KELT) survey\footnote{\url{https://keltsurvey.org}} uses two 42mm telescopes to  discover hot Jupiters orbiting bright host stars ($7<V<12$), planets well-suited for detailed atmospheric characterization \citep{Pepper:2007, Pepper:2012, Pepper:2018}. With one telescope in Sonita, AZ and the other at the South African Astronomical Observatory (SAAO) in Sutherland, South Africa, KELT surveys over 85\% of the entire sky with a 20-30 minute cadence. Each observing site has a Mamiya 645 80mm f/1.9 42mm lens with a 4k$\times$4k Apogee CCD on a Paramount ME mount. This system provides a  $26\arcdeg\times26\arcdeg$ field of view with a $23\arcsec$ pixel scale. KELT has made a significant impact on our understanding of exoplanets around early-type stars, with the discovery of 5 transiting hot Jupiters orbiting A-stars \citep{Zhou:2016, Gaudi:2017, Lund:2017, Johnson:2018, Siverd:2018} and 6 orbiting F-stars \citep{Pepper:2013, Collins:2014, Bieryla:2015, McLeod:2017,Stevens:2017, Temple:2017}. 

The planetary companion orbiting HD 93148 (hereafter \thisplanet) was identified from a joint analysis of five separate KELT-North fields that cover the celestial Northern polar cap, KN25 through KN29 (although \thisstar\ was only observed in two of the five fields). We reduced each of these KELT-North fields separately following the normal reduction process described in \citet{Siverd:2012} and \citet{Kuhn:2016}. Once the raw light curves from each field were detrended, using the trend filtering algorithm (TFA, \citealp{Kovacs:2005}), we cross-matched each field to the Tycho-2 catalog \citep{Hog:2000}. We then cross-matched the Tycho-2 IDs between the five polar cap fields from KELT-North and combined the detrended light curves into one per Tycho star. We then follow our normal candidate selection process on these combined light curves to identify a list of new polar cap candidates. We also examined the All-Sky Automated Survey for SuperNovae (ASAS-SN, \citealp{Shappee:2014, Kochanek:2017,Jayasinghe:2018}) light curves of stars nearby the KELT transit candidates to exclude nearby eclipsing binaries. \thisstar\ is located at J2000 $\alpha =$ 10$^{h}$ 47$^{m}$ 38$\fs$35101 $\delta =$ +71$\degr$ 39$\arcmin$ 21$\arcsec$15672 \citep{Gaia:2018}. \thisstar was observed 10181 times across the two KELT-North fields KN26 and KN27 from UT 2013 September 24 until UT 2017 December 31, after outliers were removed from our normal data reduction process. From our candidate selection process, we identified a candidate planet with a 5.551477 day period and a transit depth of 0.71\%. See Figure \ref{fig:DiscoveryLC} for the discovery light curve of \thisplanet.

\begin{figure}[!ht]
\vspace{0.3in}
\centering\includegraphics[width=1.0\linewidth, trim = 1.7in 0.1in 0.5in 0.1in, clip]{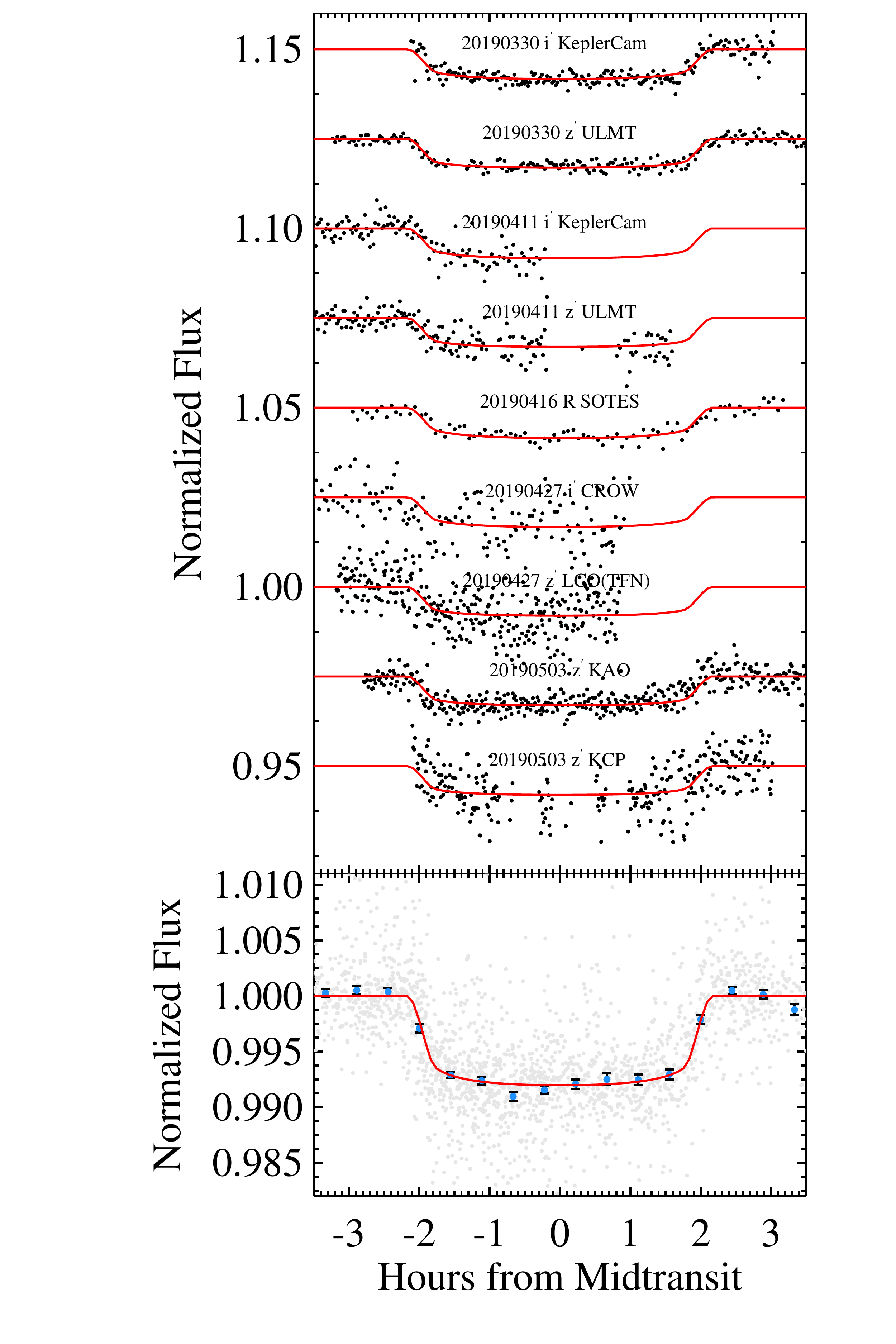}  
\caption{ {\it Top}: The KELT-FUN light curves of \thisplanet phased to global fit determined ephemeris shown in Table \ref{tab:exofast_planetary}. See Table \ref{tbl:detrending_parameters} for the information on each KELT FUN observation. The relative flux points for each observation are shown in black and the EXOFASTv2 model is plotted in red. {\it Bottom}: All light curves combined and binned to 24 minutes (blue dots with black error bars.). This combined light curve is not used in our analysis.}
\label{fig:All_LCs}
\end{figure}

\subsection{Ground-based Photometry from the KELT Follow-up Network}
\label{sec:KFUN}
Unfortunately, systematic noise and astrophysical scenarios can mimic transit signals. To rule out nearby blended eclipsing binaries and precisely measure the depth, duration, and ephemeris, we obtained multiband photometric follow-up of \thisplanet from the KELT Follow-Up Network (KELT-FUN, \citealp{Collins:2018}). KELT-FUN is a worldwide network of amateur astronomers, small-college observatories, and observing time on the Las Cumbres Observatory telescope network \citep{Brown:2013}. The telescopes range from 0.2 -- 2 meters in diameter, and this network has been responsible for the confirmation of dozens of giant planets, and the vetting of thousands of candidates. We also observed a transit of \thisplanet on UT 2019 May 03 from the Koyama Astronomical Observatory (KAO) located at Kyoto Sangyo University in Kyoto, Japan and from the Kawabe Cosmic Park (KCP) observatory in Wakayama, Japan. We used the \texttt{TAPIR} software package \citep{Jensen:2013} to schedule the observations of \thisstar. Most of the follow-up photometry was reduced and analyzed using the \texttt{AstroImageJ} astronomical observation analysis software \citep{Collins:2017}. For information on the follow-up facilities that observed KELT-24b, see Table \ref{tbl:detrending_parameters}. The follow-up transits of \thisplanet are shown in Figure \ref{fig:All_LCs}. 


\begin{figure}
	\centering\vspace{.0in}
	\includegraphics[width=1\linewidth, trim={2.5cm 13cm 8.5cm 8cm}, clip]{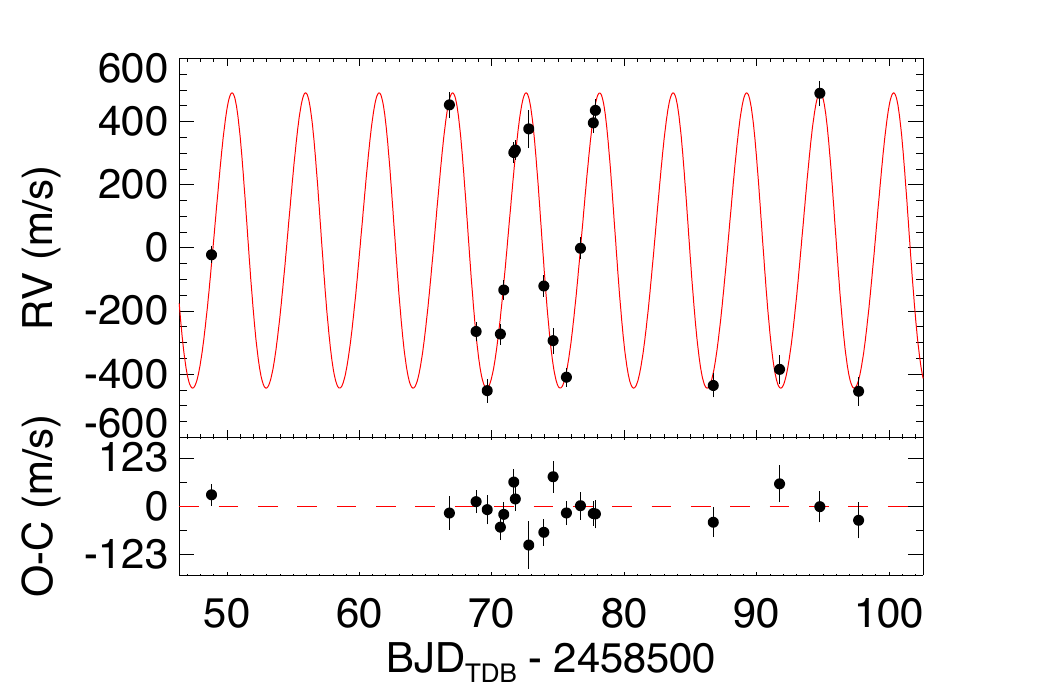}
	\includegraphics[width=1\linewidth, trim={2.5cm 13cm 8.5cm 8cm}, clip]{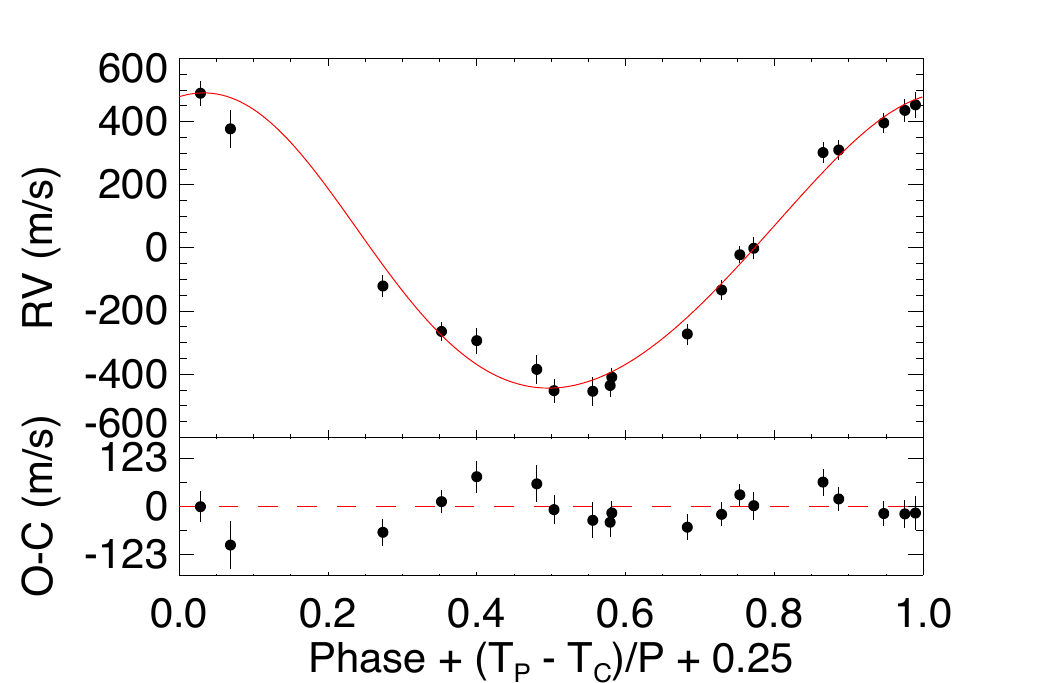}
	\caption{(Top) Radial velocity measurements from TRES (black). (Bottom) The radial velocity measurements are phase-folded to the best determined period by EXOFASTv2, 5.55 days. The EXOFASTv2 model is shown in red and the residuals to the best-fit are shown below each plot. We see no periodicity in the residuals from our fit.}
	\label{fig:TRES_RVs} 
\end{figure}

\begin{deluxetable}{l l l l l}[bt]
\tabletypesize{\small}
\tablecaption{Relative Out of Transit Radial Velocities for \thisstar from TRES \label{tab:rv}}
\tablewidth{0pt}
\tablehead{
\colhead{\bjdtdb} & \colhead{RV (m s$^{-1}$)} & \colhead{$\sigma_{RV}$ (m s$^{-1}$)} & \colhead{Bisectors} &\colhead{$\sigma_{Bis}$}}
\startdata
2458548.824216 & 402.7 & 20.4 & -8.7 & 64.5\\
2458566.788751 & 877.4 & 38.1 & -33.4 & 60.1\\
2458568.804163 & 159.8 & 22.8 & -33.2 & 72.5\\
2458569.643873 & -27.6 & 32.9 & -86.8 & 66.0\\
2458570.638165 & 151.7 & 27.7 & 197.2 & 77.6\\
2458570.893951 & 291.2 & 24.4 & -23.5 & 56.4\\
2458571.650562 & 726.4 & 29.2 & -14.0 & 45.0\\
2458571.768010 & 734.4 & 25.4 & -127.6 & 60.3\\
2458572.781102 & 801.4 & 58.1 & 37.8 & 116.0\\
2458573.918608 & 303.8 & 28.7 & 89.0 & 42.4\\
2458574.618596 & 130.8 & 36.5 & 14.4 & 63.8\\
2458575.626432 & 15.1 & 24.5 & 9.8 & 55.6\\
2458576.685400 & 423.3 & 30.4 & 9.9 & 53.6\\
2458577.655622 & 820.4 & 25.7 & -95.1 & 55.8\\
2458577.812743 & 860.0 & 31.6 & -56.8 & 46.7\\
2458586.716848 & -11.2 & 33.2 & 100.8 & 94.4\\
2458591.721299 & 39.8 & 43.2 &- 199.2 & 108.3\\
2458594.763412 & 914.2 & 35.9 & -112.1 & 63.8\\
2458597.690503 & -29.5 & 42.4 & -84.6 & 58.3\\
\hline
\enddata
\end{deluxetable}

\subsection{TRES Spectroscopy} \label{sec:TRES}
To confirm the planetary nature of \thisplanet, we obtained 59 spectra using the Tillinghast Reflector Echelle Spectrograph \citep[TRES;][]{furesz:2008}\footnote{\url{http://www.sao.arizona.edu/html/FLWO/60/TRES/GABORthesis.pdf}} on the 1.5m Tillinghast Reflector located at the Fred L. Whipple Observatory (FLWO) on Mt. Hopkins, AZ. TRES has a resolving power of R$\sim$44,000, and has been highly successful in confirming exoplanet candidates from both ground- and space-based transit surveys. We reduced the TRES spectra and extract radial velocities (RVs) following the procedure described in \citet{Buchhave:2010} and \citet{Quinn:2012} with the exception of the creation of the template spectrum used. To create a high signal-to-noise ratio template spectrum, we shifted and median-combined the out-of-transit spectra. We then used the median template to remove cosmic rays and replaced them with the appropriate section of the stellar spectrum rather than interpolating across the masked outliers. We cross-correlated the cleaned observed spectra against the median template to determine our final relative RVs (see Table \ref{tab:rv} and Figure \ref{fig:TRES_RVs}). We report RVs derived from only 19 of the 59 spectra in our orbital solution. Most of the excluded spectra were taken in-transit, for which the Rossiter-McLaughlin effect will systematically bias the RVs. We also reject all but one out-of-transit RVs from the night of the transit observation, because inclusion of all of those RVs could bias the orbital solution. That is, in the presence of stellar activity on timescales longer than the sequence of out-of-transit spectra on that night, the formal uncertainty could end up being much smaller than the systematics induced by stellar activity. While including only one RV from that night does not eliminate the possibility that stellar activity can affect the orbital solution, it does prevent an outsized effect from a single epoch. We calculated bisector spans for the 19 TRES spectra contributing to the orbital solution following the method described in \citet{Torres:2007}. We see no significant correlation between the bisector spans and the RVs. We also see no large scatter above the RV uncertainties, which are small relative to the RV semi-amplitude.

To constrain the stellar parameters \teff\ and \feh for our global analysis, we analyzed the TRES spectra using the Stellar Parameter Classification (SPC) package \citep{Buchhave:2012}. We determined the effective temperature, metallicity, surface gravity, and rotational velocity of \thisstar to be: $\teff$ = 6499 $\pm$ 50 K, $\loggstar$ = 4.28 $\pm$ 0.10, and \feh = 0.16 $\pm$ 0.08. We measure $\vsini$  = 19.46 $\pm$ 0.18 km s$^{-1}$ and a macroturbulent broadening of 10.47 $\pm$ 1.47 \kms for \thisstar\ following the method presented in \citet{Zhou:2016a} and \citet{Zhou:2018}.

\begin{figure*}[!ht]
\vspace{.0in}
\centering\includegraphics[width=.45\linewidth]{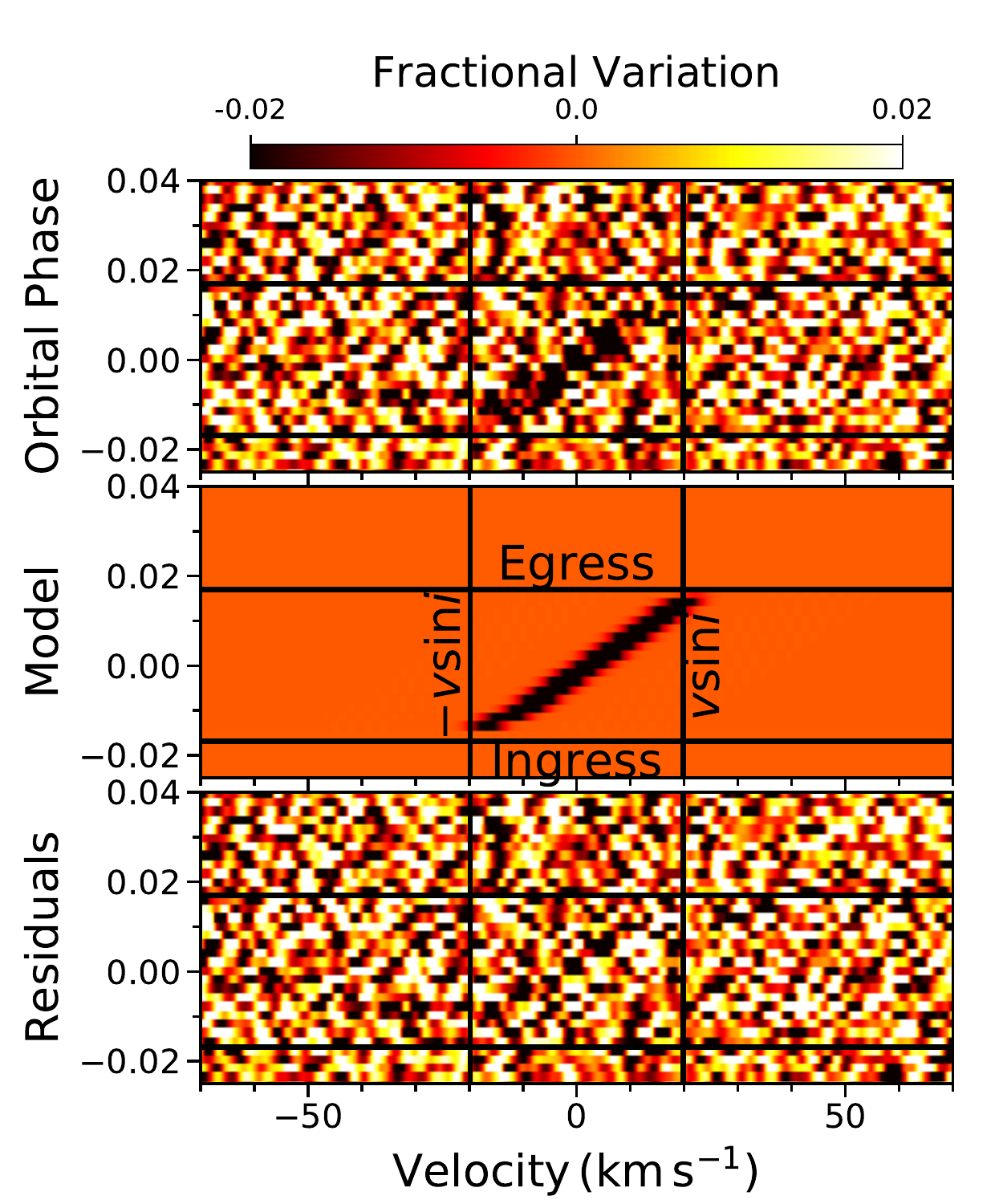}\includegraphics[width=.45\linewidth, angle = 0, trim = 0 0 0 0]{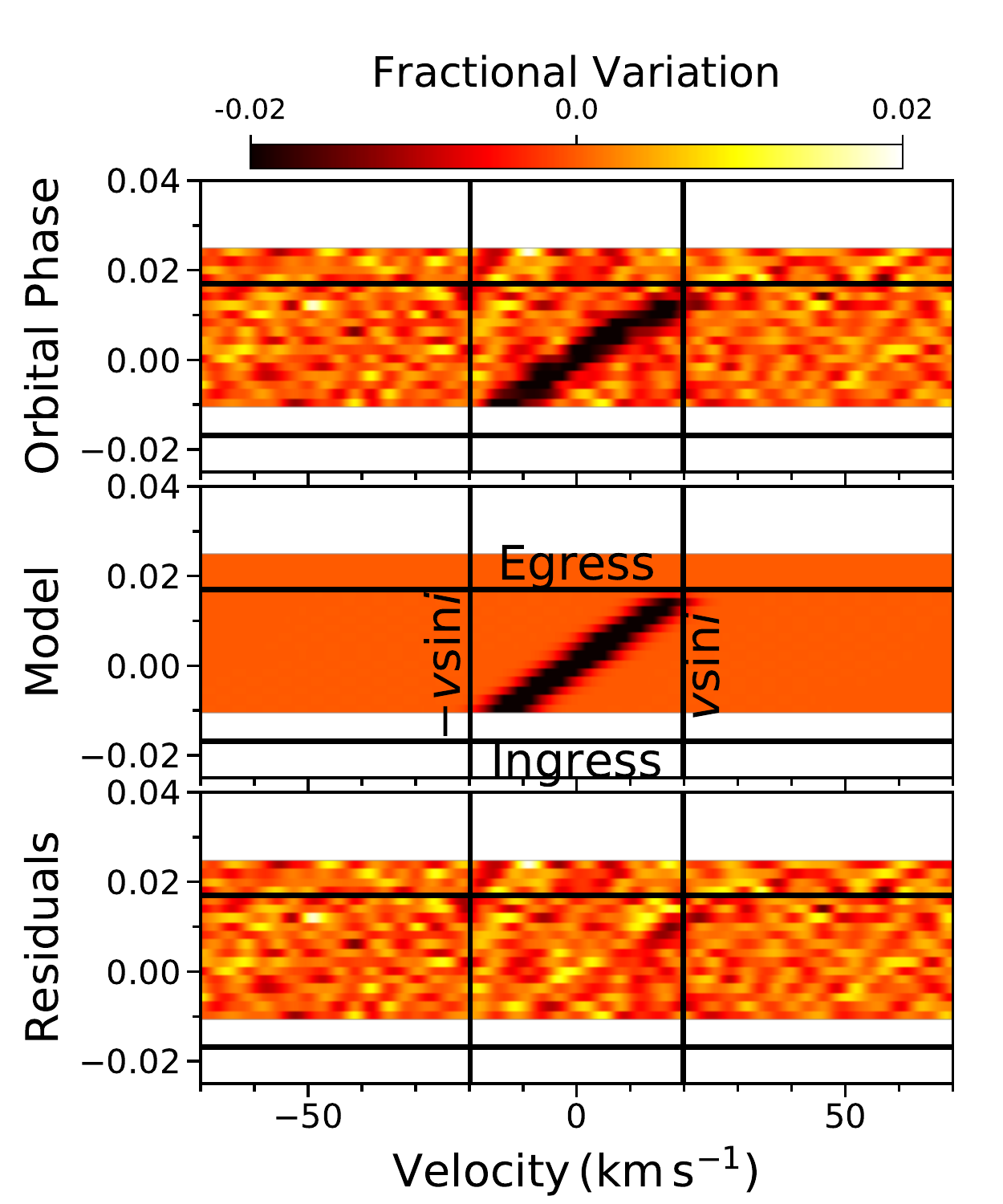}
\caption{The Doppler Tomographic transit of \thisplanet on UT 2018 March 31 from (Left) MINERVA and (Right) TRES. The overall average of the line profiles from each instrument are plotted as a function of orbital phase and velocity. The MINERVA observation is the combined Doppler tomographic signal from telescopes 2 and 3 (see \S\ref{sec:MINERVA}). The top panel in each plot represents the observed transit. The middle and bottom panels show the best-fit models and residuals after the model has been subtracted. MINERVA observed the entire transit, with baseline on each side, while TRES did not observe the beginning of the transit.
}
\label{fig:DT} 
\end{figure*}

Of the 59 TRES spectra, 40 were taken during and immediately after the transit of \thisplanet on UT 2019 March 31 with the aim of measuring the spectroscopic transit of the planet. The exposures during transit achieved a signal-to-noise ratio of 70-90 per resolution element on the Mg b lines (5187 \AA). During the transit, the planet successively blocks different parts of the rotating stellar disk, that is seen as an indentation on the spectroscopic line profile. By extracting this indentation from the stellar line profile of each spectrum, we can reveal the spectroscopic transit of the planet, a technique known as Doppler tomography \citep{CollierCameron:2010}. The Doppler tomographic (DT) signal of \thisplanet was extracted from these spectra following the methodology from \citet{Zhou:2016a}. We fit the DT signal from TRES within our global fit (see \S\ref{sec:GlobalModel} and Figure \ref{fig:DT}) to constrain the spin-orbit alignment of \thisplanet.





To derive an absolute RV for \thisstar, we cross-correlated each TRES spectrum against the CfA library of synthetic spectra \citep[see, e.g.,][]{Nordstroem:1994,Latham:2002}, which employ Kurucz model atmospheres \citep{Kurucz:1992}. The instrumental zero-point is calculated using RV standard stars that are monitored nightly and placed on the absolute RV scale from \citet{Nidever:2002}. This results in an absolute velocity of the system barycenter of $-5.749\pm0.065$ \kms. 

\subsection{MINERVA Spectroscopy} \label{sec:MINERVA}
We also obtained 37 1800-second spectroscopic exposures of \thisstar using the MINiature Exoplanet Radial Velocity Array (MINERVA) during the entire night beginning UT 2019 March 31, of which 17 exposures were taken during the transit of \thisplanet. MINERVA is an array of four PlaneWave CDK700 0.7m telescopes located at Mt. Hopkins, Arizona \citep{Swift:2015,Wilson:2019}. The four telescopes simultaneously fiber feed an R=80,000 KiwiSpec spectrograph \citep{barnes:2012,gibson:2012}, so each exposure contains a spectrum from the four telescopes, each covering roughly 500-630 nm. While MINERVA is typically calibrated with an iodine cell, we removed it during these exposures to increase throughput. \citet{Wilson:2019} showed that the vacuum-stabilized, temperature-controlled spectrograph is stable on $\sim$year-long timescales, and so we did not expect significant variation of the spectrograph during the night. An approximate wavelength solution for the DT analysis was derived from archival thorium argon exposures.

Only two of the four telescopes showed a significant signal. This was the first attempt at guiding all night on the same target, and the star drifted off the fiber due to flexure between the fiber and the guide camera in the other two telescopes before the transit began, and so their data were not used in this analysis. The DT signal was extracted from the MINERVA in-transit spectra following the technique shown by \citet{Zhou:2016a}. We simultaneously fit the Doppler tomographic signal observed from each MINERVA telescope (see \S\ref{sec:GlobalModel} and Figure \ref{fig:DT} which shows the combined MINERVA DT signal for both telescopes.) 

\subsection{Keck/NIRC2 AO Imaging}
\label{sec:AO}

\begin{figure}[!ht]
\centering 
\includegraphics[width=\linewidth,trim={1.7in 1in 0.5in 0.5in} ]{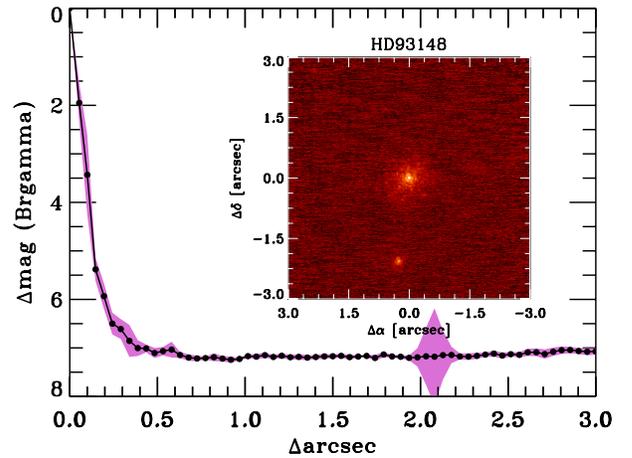}
\caption{ The Br$\gamma$-band AO image from KECK NIRC2 and 5$\sigma$ contrast curve for \thisstar. The purple swath represents the uncertainty on the 5$\sigma$ contrast curve (see \S\ref{sec:AO}).}
\label{fig:AO}
\end{figure}

The follow-up photometric observations from KELT-FUN of \thisstar\ can only detect bright nearby companions at a separation of a few arcseconds. Unfortunately, nearby unresolved companions can significantly influence the estimated planetary radius by diluting the transit depth \citep{Ciardi:2015}. Therefore, to properly account for any photometric contamination from any unaccounted stellar sources, we observed \thisstar\ with the Near Infrared Camera 2 (NIRC2) adaptive optics (AO) set up on the W. M. Keck Observatory on UT 2019 May 12 in the Br-$\gamma$ band (see Figure \ref{fig:AO}). Since \thisstar\ is very bright ($K$ = 7.154), we chose the narrower Br-$\gamma$ filter instead of the $K$-band. NIRC2 on KECK has a 1024$\times$1024 CCD and 9.942 mas pix$^{-1}$ pixel scale. Part of the detector (the lower left quadrant) suffers from higher than typical noise levels compared to the other quadrants. A 3-point dither pattern was used to avoid this part of the detector. After sky removal and flat-fielding corrections were applied, the observations of \thisstar\ were aligned and co-added to create the final image seen in Figure \ref{fig:AO}, and a final 5$\sigma$ sensitivity curve as a function of spatial separation as shown embedded in the plot. We detected a nearby star in Br-$\gamma$ with a contrast of 2.6 mag in the KECK NIRC2 AO images. {\it Gaia} detected the same star with a $\Delta G$ of 4.76 and a separation of 2.064$\pm$0.001$\arcsec$ \citep{Gaia:2018}. This star has a parallax of 11.108$\pm$0.127 mas corresponding to a distance of 90.25$\pm$1.03 pc, with a correction applied from \citealp{Stassun:2018}, and proper motions of $\mu_{\alpha},\mu_{\delta}=-50.756 \pm 0.325, -37.811 \pm 0.200~{\rm mas~yr}^{-1}$. These proper motions are significantly different from the proper motions of \thisstar: $\mu_{\alpha},\mu_{\delta}=-56.184\pm0.053 , -34.808\pm0.064~{\rm mas~yr}^{-1}$. The difference in proper motion could be explained by the orbital motion of the nearby companion to \thisstar but the estimated radial distances to each star from {\it Gaia} differ by 5.7 pc. Since the two stars only have a projected separation of 2.064$\arcsec$ (186 au), they are physically separated by $\sim$5.7 pc. Therefore, it is not clear whether this companion is bound to \thisstar.

We determined the sensitivity to any additional nearby bound or unbound companions by injecting simulated sources with a S/N of 5 azimuthally around the primary target every $45^\circ$ at separations of integer multiples of the central source's FWHM. The contrast limits at each injected location were determined from the brightness of the injected sources relative to \thisstar. We average all of the determined limits at each radial separation to establish the 5$\sigma$ detection limit at that distance. The rms dispersion of these azimuthally averaged limits set the uncertainty at each radial separation \citep{Furlan:2017}. 

The nearby faint companion is blended in all of our photometric follow-up observations from KELT-FUN. To create the $\sim$0.7\% transit seen in our follow up photometry from KELT-FUN, this companion would need to have a $\sim$59\% deep eclipse. While unlikely, it is not impossible for an eclipsing binary to have this deep of an eclipse. However, due to its small flux contribution to the spectroscopic line profiles, a companion this faint is not able to significantly influence the RVs. It is possible that a faint companion can slightly affect the measured RVs, but this would only be at the level of a few m/s, not the hundreds of m/s we detect for the orbit of \thisplanet (K = 458 \ms, see discussion of blended CCFs in \citealp{Buchhave:2011}). Therefore, our subsequent RV follow up confirms that the planetary companion is orbiting our target star (\thisstar) and not the faint companion detected in {\it Gaia} and our KECK AO observations. To properly determine the size of \thisplanet, we account for the companion's contribution in our global analysis (see \S\ref{sec:GlobalModel}).


\subsection{SED Analysis}
\label{sec:SED}
\begin{figure}[!ht]
\centering 
\includegraphics[trim = 1.3in 0.6in 1in 1.15in , width=\columnwidth]{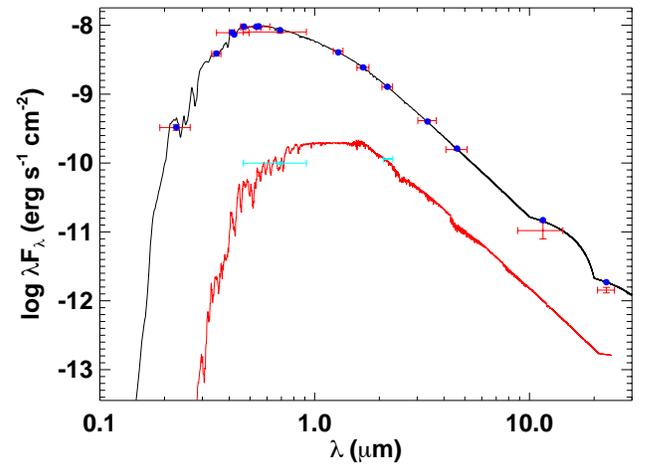}
\caption{The two-component SED fit for \thisstar and its companion. The blue points are the predicted integrated fluxes and the red points are the observed values at the corresponding passbands. The width of the bandpasses are the horizontal red error bars and the vertical errors represent the 1$\sigma$ uncertainties. The cyan points are the $G$ and Br-$\gamma$ fluxes from {\it Gaia} and our AO observations (see \S\ref{sec:AO}). The best-fit atmospheric model for \thisstar\ is shown by the black solid and the companion is in red. }
\label{fig:sed_fit}
\end{figure}

The spectral energy distribution (SED) was not included within our global fit due to the presence of the nearby companion seen by {\it Gaia} and our AO observations (see \S\ref{sec:AO} and Figure \ref{fig:AO}). Therefore, we performed a 2-component SED fit to determine the flux contribution from the companion in each of our follow-up photometric filters to use as an input into our global analysis (see \S\ref{sec:GlobalModel}). Using the available broadband photometry shown in Table \ref{tbl:LitProps}, we fit the SED for \thisstar\ from 0.2--20~$\mu$m (Figure~\ref{fig:sed_fit}). We applied a minimum uncertainty to the reported errors to account for a systematic error floor in the broadband magnitudes reported in the various catalogs. The nearby companion has a $\Delta$G of 4.76 and a $\Delta$Br-$\gamma$ of 2.6 mag. We assumed both stars have the same extinction ($A_V$) and use the observed $\Delta G$ and $\Delta$Br-$\gamma$ to fit an SED for the faint companion (see Figure \ref{fig:sed_fit}). The \citet{Kurucz:1992} stellar atmosphere models were used to fit each flux point for the primary and the NextGen model atmosphere grid were used for the companion \citep{Hauschildt:1999}, and we use the SPC determined \teff\ and \feh\ as Gaussian priors. We also used the \loggstar\ from the global fit (see \S\ref{sec:GlobalModel}) as a Gaussian prior. We allowed $A_V$ to be a free parameter but constrain it to the maximum permitted line-of-sight extinction from \citet{Schlegel:1998}. 

The final SED fit has a reduced $\chi^2$ of 2.7, an extinction of $A_V = 0.11 \pm 0.02$ mag (see Figure \ref{fig:sed_fit}), and an unextincted bolometric flux received at Earth of $F_{\rm bol} = 1.309\pm 0.015 \times 10^{-10}$ erg~s$^{-1}$~cm$^{-2}$ (correcting for the contamination of the companion). We combined the bolometric flux with the \teff\ from our SPC analysis (which was adopted for this fit) to measure the radius of \thisstar to be $R_\star = 1.526\pm 0.022$~\rsun. We enforced a Gaussian prior on \rstar\ in our global fit (see \S\ref{sec:GlobalModel}). The flux contribution from the nearby companion is 1.01\% ($R$),  1.79\% (i$^{\prime}$),  and 2.73\% (z$^{\prime}$).


\subsection{Location in the Galaxy, UVW Space Motion, and Galactic Population}
\label{sec:uvw}
We determined the three-dimensional Galactic space motion of \thisstar\ to understand its location within the Milky Way galaxy and the Galactic population it belongs to. \thisstar\ is located at $\alpha_{\rm J2000} =10^h47^m38\fs351$ and $\delta_{\rm J2000} =+71\arcdeg39\arcmin21\farcs157$, and from {\it Gaia} DR2 the parallax is $10.414\pm0.0469$~mas (after applying the correction from \citealp{Stassun:2018}). Ignoring the Lutz-Kelker bias, which should be negligible \citep{Lutz:1973}, this star is located at a distance of 96.02$\pm$0.43 pc from the Sun. Combining the sky position and distance, \thisstar\ is located at a vertical (Z-Z$_{\odot}$) distance of 64.6 pc from the Sun.  \citet{Bovy:2017} estimates from \textit{Gaia} that the Sun is located at a vertical distance above the plane of Z$_{\odot} \sim 30$ pc. Therefore, \thisstar\ is located at Z$\sim$100 pc above the plane.  This is the typical scale height for mid-to-late F thin disk stars \citep{Bovy:2017}. Using the \textit{Gaia} parallax and proper motions $\left(\mu_{\alpha},\mu_{\delta})=(-56.184 \pm 0.053, -34.808 \pm 0.064~{\rm mas~yr}^{-1} \right)$ and the absolute radial velocity as determined from the TRES spectroscopy of $-5.749\pm0.065  ~{\rm km~s^{-1}}$, we calculated the three-dimensional Galactic space motion of $(U,V,W)=(-11.00 \pm 0.11, -9.36 \pm 0.10, 0.11\pm 0.05)~{\rm km~s^{-1}}$, where positive $U$ is in the direction of the Galactic center and adopting the \citet{Coskunoglu:2011} determination of the solar motion with respect to the local standard of rest. The relatively low W velocity of KELT-24 suggests that KELT 24 may be close to its maximum excursion above the plane. \thisstar\ has a 99.5\% chance of being located in the thin disk, according to the classification of \citet{Bensby:2003}. The location of KELT-24 and its relatively low UVW velocities are both consistent with it being a young star, which corroborates the relatively young age inferred from evolutionary models (see Figure \ref{fig:hrd}). The only association that is close to the estimated distance (96 pc) and UVW velocities of \thisstar\ is the extended Ursa Major moving group. While the distance and 3-D space motion of \thisstar\ are clearly inconsistent with the core of the association ($\sim$24 pc), its distance is consistent with known members of the Ursa Major moving group stream ($\sim$100pc). However, a full UVW analysis of the entire association using the Gaia DR2 proper motions and distances are needed to conclusively determine whether \thisstar\ is a member of Ursa Major association. A detailed analysis of whether or not KELT-24 is a member of the Ursa Major association is well outside of the scope of this paper, but we advocate that this is a worthwhile exercise, particularly given the other evidence for the youth of the host star presented in this paper.

\begin{table}
\scriptsize
\centering
\caption{Median values and 68\% confidence interval for global model of \thisstar}
\begin{tabular}{llcccc}
  \hline
  \hline
Parameter & Units & Values & & & \\
\hline
\multicolumn{2}{l}{Stellar Parameters:}&\smallskip\\
~~~~$M_*$\dotfill &Mass (\msun)\dotfill &$1.460^{+0.055}_{-0.059}$\\
~~~~$R_*$\dotfill &Radius (\rsun)\dotfill &$1.506\pm0.022$\\
~~~~$L_*$\dotfill &Luminosity (\lsun)\dotfill &$3.67^{+0.16}_{-0.15}$\\
~~~~$\rho_*$\dotfill &Density (cgs)\dotfill &$0.603^{+0.032}_{-0.033}$\\
~~~~$\log{g}$\dotfill &Surface gravity (cgs)\dotfill &$4.247^{+0.019}_{-0.021}$\\
~~~~$T_{\rm eff}$\dotfill &Effective Temperature (K)\dotfill &$6509^{+50}_{-49}$\\
~~~~$[{\rm Fe/H}]$\dotfill &Metallicity (dex)\dotfill &$0.186^{+0.077}_{-0.076}$\\
~~~~$[{\rm Fe/H}]_{0}^\dagger$\dotfill &Initial Metallicity \dotfill &$0.284^{+0.059}_{-0.058}$\\
~~~~$Age$\dotfill &Age (Gyr)\dotfill &$0.78^{+0.61}_{-0.42}$\\
~~~~$EEP^\ddagger$\dotfill &Equal Evolutionary Point \dotfill &$323^{+16}_{-23}$\\
~~~~$vsinI_*$\dotfill &Projected rotational velocity (km/s)\dotfill &$19.76\pm0.160$\\
~~~~$\xi$\dotfill &Macroturbulence velocity (km/s)\dotfill &$5.76^{+0.51}_{-0.50}$\\
\hline
\end{tabular}
\begin{flushleft} 
  \footnotesize{ 
    \textbf{\textsc{NOTES:}}
$^\dagger$The initial metallicity is the metallicity of the star when it was formed.\\
$^\ddagger$The Equal Evolutionary Point corresponds to static points in a stars evolutionary history when using the MIST isochrones and can be a proxy for age. See \S2 in \citet{Dotter:2016} for a more detailed description of EEP.
               }
 \end{flushleft}
\label{tab:exofast_stellar}
\end{table}
\begin{table*}
\scriptsize
\centering
\caption{Median values and 68\% confidence interval for global model of \thisstar}
\begin{tabular}{llcccc}
  \hline
  \hline
Parameter & Description (Units) & Values & & & \\
\hline
~~~~$P$\dotfill &Period (days)\dotfill &$5.5514926^{+0.0000081}_{-0.0000080}$\\
~~~~$R_P$\dotfill &Radius (\rj)\dotfill &$1.272\pm0.021$\\
~~~~$T_C$\dotfill &Time of conjunction (\bjdtdb)\dotfill &$2457147.0529^{+0.0020}_{-0.0021}$\\
~~~~$T_0^\dagger$\dotfill &Optimal conjunction Time (\bjdtdb)\dotfill &$2458540.47759^{+0.00036}_{-0.00035}$\\
~~~~$a$\dotfill &Semi-major axis (AU)\dotfill &$0.06969^{+0.00087}_{-0.00096}$\\
~~~~$i$\dotfill &Inclination (Degrees)\dotfill &$89.17^{+0.59}_{-0.75}$\\
~~~~$e$\dotfill &Eccentricity \dotfill &$0.077^{+0.024}_{-0.025}$\\
~~~~$\omega_*$\dotfill &Argument of Periastron (Degrees)\dotfill &$55^{+13}_{-15}$\\
~~~~$T_{eq}$\dotfill &Equilibrium temperature (K)\dotfill &$1459\pm16$\\
~~~~$M_P$\dotfill &Mass (\mj)\dotfill &$5.18^{+0.21}_{-0.22}$\\
~~~~$K$\dotfill &RV semi-amplitude (m/s)\dotfill &$462^{+16}_{-15}$\\
~~~~$logK$\dotfill &Log of RV semi-amplitude \dotfill &$2.665\pm0.015$\\
~~~~$R_P/R_*$\dotfill &Radius of planet in stellar radii \dotfill &$0.08677^{+0.00071}_{-0.00070}$\\
~~~~$a/R_*$\dotfill &Semi-major axis in stellar radii \dotfill &$9.95^{+0.17}_{-0.18}$\\
~~~~$\delta$\dotfill &Transit depth (fraction)\dotfill &$0.00753\pm0.00012$\\
~~~~$Depth$\dotfill &Flux decrement at mid transit \dotfill &$0.00753\pm0.00012$\\
~~~~$\tau$\dotfill &Ingress/egress transit duration (days)\dotfill &$0.01458^{+0.00084}_{-0.00029}$\\
~~~~$T_{14}$\dotfill &Total transit duration (days)\dotfill &$0.17917^{+0.0011}_{-0.00097}$\\
~~~~$T_{FWHM}$\dotfill &FWHM transit duration (days)\dotfill &$0.16442^{+0.00081}_{-0.00080}$\\
~~~~$b$\dotfill &Transit Impact parameter \dotfill &$0.134^{+0.13}_{-0.096}$\\
~~~~$b_S$\dotfill &Eclipse impact parameter \dotfill &$0.15^{+0.13}_{-0.11}$\\
~~~~$\tau_S$\dotfill &Ingress/egress eclipse duration (days)\dotfill &$0.01677^{+0.00068}_{-0.00062}$\\
~~~~$T_{S,14}$\dotfill &Total eclipse duration (days)\dotfill &$0.2028^{+0.0096}_{-0.010}$\\
~~~~$T_{S,FWHM}$\dotfill &FWHM eclipse duration (days)\dotfill &$0.1861^{+0.0090}_{-0.0100}$\\
~~~~$\delta_{S,3.6\mu m}$\dotfill &Blackbody eclipse depth at 3.6$\mu$m (ppm)\dotfill &$431\pm14$\\
~~~~$\delta_{S,4.5\mu m}$\dotfill &Blackbody eclipse depth at 4.5$\mu$m (ppm)\dotfill &$599\pm17$\\
~~~~$\rho_P$\dotfill &Density (cgs)\dotfill &$3.13\pm0.19$\\
~~~~$logg_P$\dotfill &Surface gravity \dotfill &$3.900^{+0.021}_{-0.022}$\\
~~~~$\lambda$\dotfill &Projected Spin-orbit alignment (Degrees)\dotfill &$2.6^{+5.1}_{-3.6}$\\
~~~~$\Theta$\dotfill &Safronov Number \dotfill &$0.389\pm0.014$\\
~~~~$\fave$\dotfill &Incident Flux (\fluxcgs)\dotfill &$1.022^{+0.043}_{-0.042}$\\
~~~~$T_P$\dotfill &Time of Periastron (\bjdtdb)\dotfill &$2457146.60^{+0.17}_{-0.22}$\\
~~~~$T_S$\dotfill &Time of eclipse (\bjdtdb)\dotfill &$2457144.423^{+0.060}_{-0.056}$\\
~~~~$T_A$\dotfill &Time of Ascending Node (\bjdtdb)\dotfill &$2457145.842^{+0.053}_{-0.057}$\\
~~~~$T_D$\dotfill &Time of Descending Node (\bjdtdb)\dotfill &$2457148.398\pm0.048$\\
~~~~$ecos{\omega_*}$\dotfill & \dotfill &$0.041^{+0.017}_{-0.016}$\\
~~~~$esin{\omega_*}$\dotfill & \dotfill &$0.063^{+0.023}_{-0.027}$\\
~~~~$M_P\sin i$\dotfill &Minimum mass (\mj)\dotfill &$5.18^{+0.21}_{-0.22}$\\
~~~~$M_P/M_*$\dotfill &Mass ratio \dotfill &$0.00339^{+0.00013}_{-0.00012}$\\
~~~~$d/R_*$\dotfill &Separation at mid transit \dotfill &$9.32\pm0.39$\\
~~~~$P_T$\dotfill &A priori non-grazing transit prob \dotfill &$0.0980^{+0.0043}_{-0.0040}$\\
~~~~$P_{T,G}$\dotfill &A priori transit prob \dotfill &$0.1166^{+0.0051}_{-0.0047}$\\
~~~~$P_S$\dotfill &A priori non-grazing eclipse prob \dotfill &$0.08622^{+0.0020}_{-0.00084}$\\
~~~~$P_{S,G}$\dotfill &A priori eclipse prob \dotfill &$0.1026^{+0.0025}_{-0.0010}$\\
\smallskip\\\multicolumn{2}{l}{Wavelength Parameters:}&R&i'&z'\smallskip\\
~~~~$u_{1}$\dotfill &linear limb-darkening coeff \dotfill &$0.258\pm0.046$&$0.217\pm0.027$&$0.148\pm0.021$\\
~~~~$u_{2}$\dotfill &quadratic limb-darkening coeff \dotfill &$0.320^{+0.049}_{-0.048}$&$0.323\pm0.028$&$0.300\pm0.022$\\
~~~~$A_D$\dotfill &Dilution from neighboring stars \dotfill &$0.01004^{+0.00050}_{-0.00051}$&$0.01761\pm0.00088$&$0.0268\pm0.0013$\\
\smallskip\\\multicolumn{2}{l}{Telescope Parameters:}&TRES\smallskip\\
~~~~$\gamma_{\rm rel}$\dotfill &Relative RV Offset (m/s)\dotfill &$416^{+12}_{-11}$\\
~~~~$\sigma_J$\dotfill &RV Jitter (m/s)\dotfill &$33^{+16}_{-14}$\\
~~~~$\sigma_J^2$\dotfill &RV Jitter Variance \dotfill &$1150^{+1300}_{-740}$\\
\smallskip\\\multicolumn{2}{l}{Doppler Tomography Parameters:}&&&\smallskip\\
~~~~$\sigma_{DT}$\dotfill &Doppler Tomography Error scaling \dotfill &$0.9932\pm0.0095$&--&--\\
\end{tabular}
 \begin{flushleft} 
  \footnotesize{ 
    \textbf{\textsc{\hspace{0.75in}NOTES:}}
See Table 3 in \citet{Eastman:2019} for a list of the derived and fitted parameters in EXOFASTv2.\\
$^\dagger$Minimum covariance with period.
All values in this table for the secondary occultation of \thisstar\ b are predicted values from our global analysis.               
               }
 \end{flushleft}
\label{tab:exofast_planetary}
\end{table*}

\begin{table*}
\small
\centering
\caption{Median values and 68\% confidence interval for global model of \thisstar}
\begin{tabular}{llcccccc}
  \hline
  \hline
\smallskip\\\multicolumn{2}{l}{Transit Parameters:}&KeplerCam UT 2019-03-30 (i')&ULMT UT 2019-03-30 (z')&KeplerCam UT 2019-04-10 (i')\\
~~~~$\sigma^{2}$\dotfill &Added Variance \dotfill &$0.00000330^{+0.00000036}_{-0.00000032}$&$0.00000115^{+0.00000017}_{-0.00000015}$&$0.00000834^{+0.0000011}_{-0.00000097}$\\
~~~~$F_0$\dotfill &Baseline flux \dotfill &$1.00015\pm0.00027$&$1.00017\pm0.00011$&$1.00026\pm0.00058$\\
~~~~$C_{0}$\dotfill &Additive detrending coeff \dotfill &$-0.00086\pm0.00058$&$0.00019^{+0.00022}_{-0.00023}$&$0.0007\pm0.0013$\\
~~~~$C_{1}$\dotfill &Additive detrending coeff \dotfill &---&$0.00032\pm0.00022$&---\\
\smallskip\\\multicolumn{2}{l}{Transit Parameters:}&ULMT UT 2019-04-11 (z')&SOTES UT 2019-04-16 (R) &CROW UT 2019-04-27 (i')\\
~~~~$\sigma^{2}$\dotfill &Added Variance \dotfill &$0.00000680^{+0.00000085}_{-0.00000073}$&$-0.00000069^{+0.00000049}_{-0.00000040}$&$0.0000370^{+0.0000057}_{-0.0000048}$\\
~~~~$F_0$\dotfill &Baseline flux \dotfill &$0.99985\pm0.00021$&$1.00034^{+0.00041}_{-0.00040}$&$2.9437\pm0.0013$\\
~~~~$C_{0}$\dotfill &Additive detrending coeff \dotfill &---&$-0.00060\pm0.00077$&$0.00451^{+0.00085}_{-0.00084}$\\
\smallskip\\\multicolumn{2}{l}{Transit Parameters:}&LCO TFN UT 2019-04-27 (z')&KAO UT 2019-05-03 (z')& KCP UT 2019-05-03 (z')\\
~~~~$\sigma^{2}$\dotfill &Added Variance \dotfill &$0.0000333^{+0.0000034}_{-0.0000030}$&$0.00000560^{+0.00000054}_{-0.00000048}$&$0.0000260^{+0.0000028}_{-0.0000025}$\\
~~~~$F_0$\dotfill &Baseline flux \dotfill &$1.00556^{+0.00079}_{-0.00077}$&$1.00246\pm0.00029$&$0.99967\pm0.00039$\\
~~~~$C_{0}$\dotfill &Additive detrending coeff \dotfill &$-0.0000^{+0.0016}_{-0.0015}$&$0.00266^{+0.00060}_{-0.00059}$&$-0.0003\pm0.0032$\\
\hline
  \hline
\label{tab:KELT-24_other}
 \end{tabular}
\end{table*}

\section{EXOFAST\lowercase{v}2 Global Fit for \thisstar} 
\label{sec:GlobalModel}
\begin{figure}[!ht]
    \centering
    \includegraphics[width=\linewidth, trim={0.5cm 0cm 0cm 0.5cm }, clip]{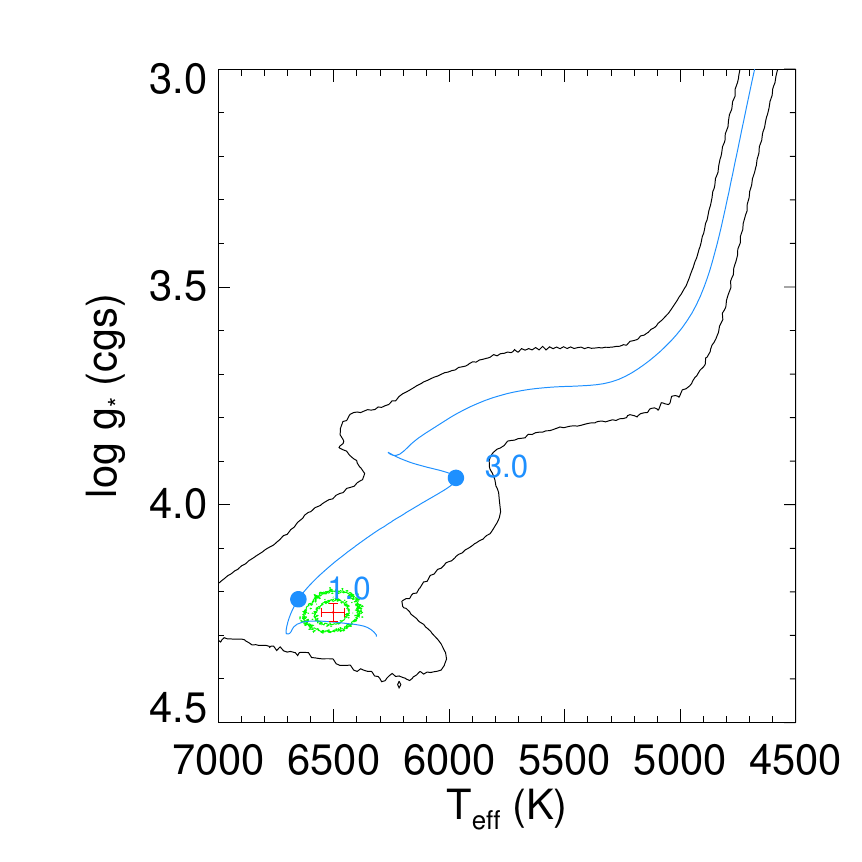}
    \caption{The best fitting MIST track is shown by the blue line. The 3$\sigma$ contours for the MIST evolutionary tracks are shown in black. The median values and 1$\sigma$ errors from our global fit for \teff\ and \feh\ are shown in red with the corresponding 3$\sigma$ contours in green. The blue points represent the 1.0 and 3.0 Gyr positions along the MIST track.}
    \label{fig:hrd}
\end{figure}

To understand the system parameters and place \thisplanet in the context of all known planets, we globally fit all available photometry and spectroscopic observations using the publicly available exoplanet modeling suite EXOFASTv2 \citep{Eastman:2013, Eastman:2017, Eastman:2019}. We simultaneously fit the transit light curves from KELT-FUN (see Table \ref{tbl:detrending_parameters} and Figure \ref{fig:All_LCs}) with the RVs from TRES (see Table \ref{tab:rv} and Figure \ref{fig:TRES_RVs}). We enforced a Gaussian prior on the ephemeris of T$_C$ = 2457147.0522$\pm$0.0021 \bjdtdb\ and P = 5.551467$\pm$0.000034 days from an EXOFASTv2 fit of just the KELT-North data. Within this analysis we also fit the DT signals observed on UT 2019 March 31 by MINERVA (two telescopes fit separately) and TRES (see Figure \ref{fig:DT}). The host star was characterized within the fit using the MESA Isochrones and Stellar Tracks (MIST) stellar evolution models \citep{Dotter:2016, Choi:2016, Paxton:2011, Paxton:2013, Paxton:2015}. The best-fit MIST evolutionary track is shown in Figure \ref{fig:hrd}. From the SPC analysis of the TRES spectra (see \S\ref{sec:TRES}), we enforced a Gaussian prior on \teff\ (6499$\pm$50 K), \feh\ (0.16$\pm$0.08), and \vsini\ (19.458$\pm$0.182 \kms). From {\it Gaia}, AO, and our 2-component SED fit, we know that the nearby companion contributes 1.01\% in $R$, 1.79\% in i$^{\prime}$, and 2.73\% in z$^{\prime}$. To properly account for this contribution in the follow-up observations, we used these flux contributions with a 5\% error as Gaussian priors in the EXOFASTv2 global fit.  We note that the dilution prior on the follow-up photometry has no influence on the determined results. We also placed a prior on the radius of \thisstar of $R_\star = 1.526\pm 0.022$~\rsun, from our 2-component SED fit. The final results from our EXOFASTv2 fit of the \thisstar\ system are shown in Tables \ref{tab:exofast_stellar}, \ref{tab:exofast_planetary}, and \ref{tab:KELT-24_other}. We refer the reader to Table 3 in \citet{Eastman:2019} for a list of the derived and fitted parameters in EXOFASTv2.


The KELT-North data has a time baseline of over 4 years, covering 64 different transits of \thisplanet. Therefore, we explored the possibility of including the KELT-North photometry in the EXOFASTv2 fit to provide a better constraint on the ephemeris of the transit for future follow-up. However, we ran tests to ensure that the lower precision of the KELT-North photometry did not significantly influence the resulting system parameters. As a result of \thisstar\ being observed in two separate fields and KELT avoiding observing within 50$^{\degr}$ of the moon, the number of observations in each transit varies significantly, with a maximum of 54 observations over a 4.9 hour transit (plus$\sim$1 hour baseline on each side). We ran two separate EXOFASTv2 fits, one as described in the previous paragraph that excluded the KELT-North data but placed a Gaussian prior on the ephemeris of \thisplanet (T$_C$ and period) from an EXOFASTv2 fit of just the KELT-North data.  We also ran another fit where we included all 64 transits from KELT-North plus the KELT-FUN follow-up transits (see Figure \ref{fig:All_LCs}). From this test, we saw no evidence that the inclusion of the KELT-North observations significantly influenced the results since the two fits were consistent to within $<1\sigma$ on all parameters. We did see a small (17.5\%) improvement on the precision of \thisplanet's period when including the KELT-North transits in the global fit. We note that this difference in precision corresponds to $<$1 second. The optimal time of conjunction had a similar precision between the two fits. We did notice that the inclusion of the KELT data resulted in a duration that is shorter (than the fit excluding the KELT data) by 25 seconds. Although this is within the 1$\sigma$ uncertainty on the transit duration from our results (see Table \ref{tab:exofast_planetary}), we choose to not include the KELT observations within the global fit as a precaution.


\begin{figure*}[!ht]
\vspace{.1in}
\centering
\includegraphics[trim = 0 6.4in 0 0 , width=0.9\linewidth]{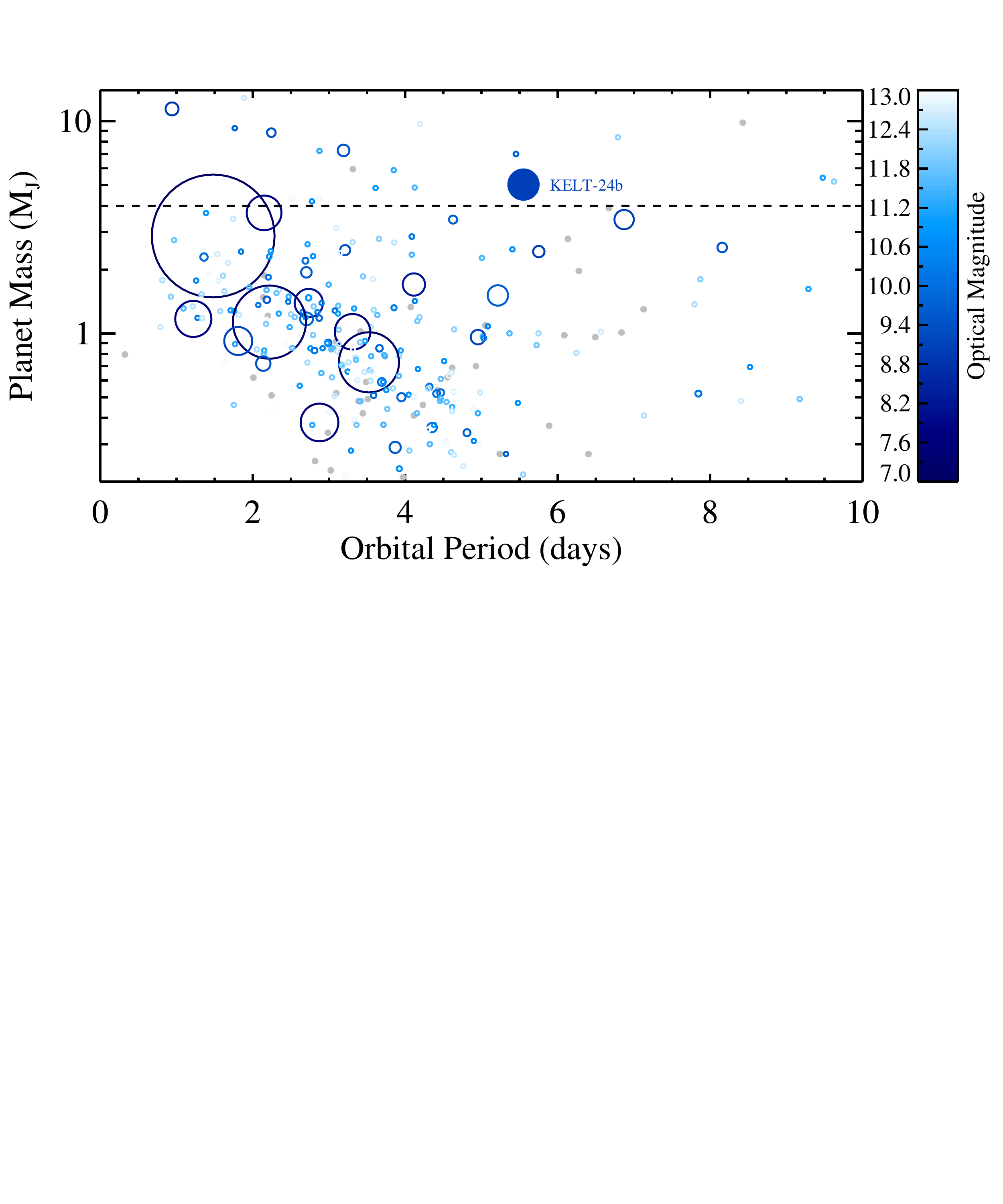}
\caption{The distribution of planet mass and orbital period for the known population of radial velocity only (gray) and transiting hot Jupiters (colored by optical magnitude). The size of the circle is scaled by the host star's apparent brightness. The filled in circle represents the location of \thisplanet. We only show systems that have a 3$\sigma$ or better measurement on the planet's mass. The horizontal dashed line is the lower limit (4 \mj) of the massive hot Jupiters regime we discuss in \S\ref{sec:discussion}. The data behind this figure was downloaded from the composite table on UT 2019 May 07 from the NASA Exoplanet Archive \citep{Akeson:2013}.
}
\label{fig:discussion} 
\end{figure*}

\section{Discussion}
\label{sec:discussion}
\thisplanet has some key characteristics that make it a compelling target for detailed characterization. Specifically, the host star is very bright, $V$ = 8.3 mag, and the planet is quite massive, \pmass\ \mj. With such a high mass, it is interesting to see some signs that it is inflated (R$_P$ = \prad\ \rj). However, this is not unique to this system since many massive hot Jupiters have inflated radii. Of all the hot Jupiters known, \thisplanet is one of only a few dozen massive (M$_P$ = 4--13 \mj) hot Jupiters (P$<$10 days) with a host star bright enough ($V$<13 mag) to permit detailed characterization.\footnote{\url{https://exoplanetarchive.ipac.caltech.edu/}; \citet{Akeson:2013};  UT 2019 May 07} At $V$ = 8.3 mag, \thisstar\ is the brightest known planetary host in this regime (see Figure \ref{fig:discussion}). The host star, \thisstar, has a mass of $M_{\star}$ = \starmass\ \msun, a radius of $R_{\star}$ = \starradius\ \rsun, and an age of \starage\ Gyr. It is the brightest star known to host a transiting giant planet with a period between 5 and 10 days, and one of the longest period planets discovered from ground-based surveys. Interestingly, HAT-P-2b \citep{Bakos:2007} is quite similar to \thisplanet\, in that they have almost the same orbital period (5.63 days compared to 5.55 days), similar planetary masses (9.0 M$_J$ compared to 5.2 M$_J$), and both host stars are very bright (HAT-P-2 is $V$ = 8.7 mag). The relatively young age of \thisstar\ suggests it has just started to evolve from the zero-age main sequence, which is consistent with our UVW analysis (see \S\ref{sec:uvw}). 

We detected a non-zero, small 3$\sigma$ eccentricity of \pecc\ for \thisplanet's orbit. However, systems observed to have small eccentricities ($<$0.1) are subject to the Lucy-Sweeney bias, where observational errors of a circular orbit can lead to the detection of a slight eccentricity \citep{Lucy:1971}. Therefore, we caution the reader about the detection of the eccentricity, even though it is detected at a formally significant confidence level. We do note that this eccentricity was not only constrained by the spectroscopic observations from TRES (see \S\ref{sec:TRES}) but also from the KELT-FUN transit observations (i.e. the transit duration), since they are all globally modelled with EXOFASTv2 (see \S\ref{sec:GlobalModel}). Since the eccentricity is quite small and not conclusive, we use equation 3 from \citet{Adams:2006} to approximate the circularization timescale of \thisplanet to be 12.7 Gyrs (assuming Q$_{\star}$ = 10$^6$). This circularization timescale does not change significantly when accounting for the small eccentricity detected. Since the age of \thisstar\ is significantly smaller than the circularization timescale, we do not assume the eccentricity to be zero within our global analysis. Future observations should confirm this non-zero eccentricity by obtaining additional higher precision radial velocities and/or observing the secondary eclipse of \thisplanet. The time difference between the secondary eclipse assuming zero eccentricity and one using e = 0.078 from our results is about 3.5 hours. Future eclipse observations should account for this when scheduling eclipse observations. \thisstar has a projected rotational velocity of 19.46 $\pm$ 0.18 km s$^{-1}$, corresponding to a rotation period of 3.9 days. Since this is shorter than the orbital period of \thisplanet\, we do not expect the planet to tidally synchronized.

\subsection{Tidal Evolution and Irradiation History}
\label{sec:evo} 
\begin{figure}[!ht]
\vspace{.1in}
\centering
\includegraphics[width=1\linewidth]{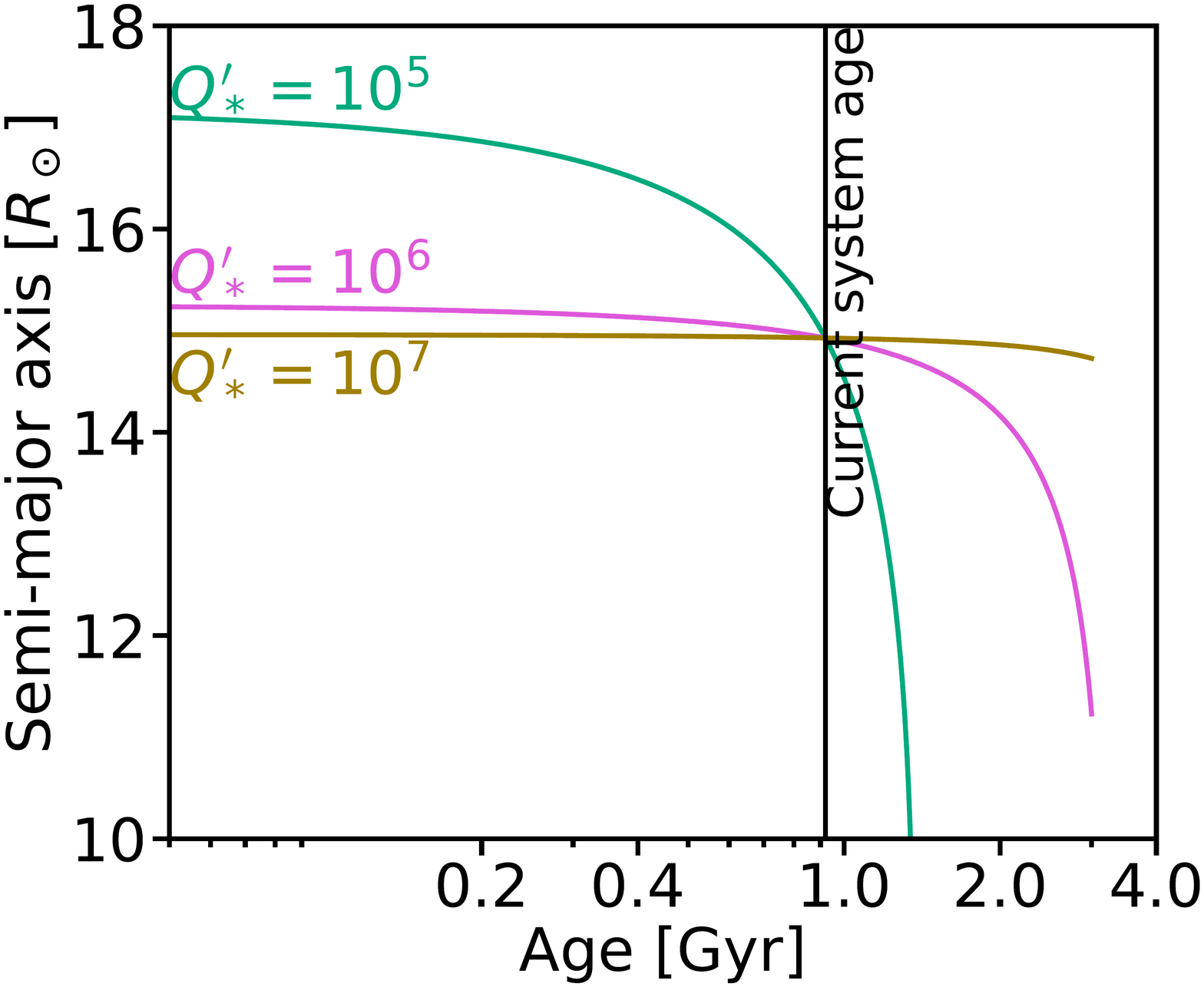}

\includegraphics[width=1\linewidth]{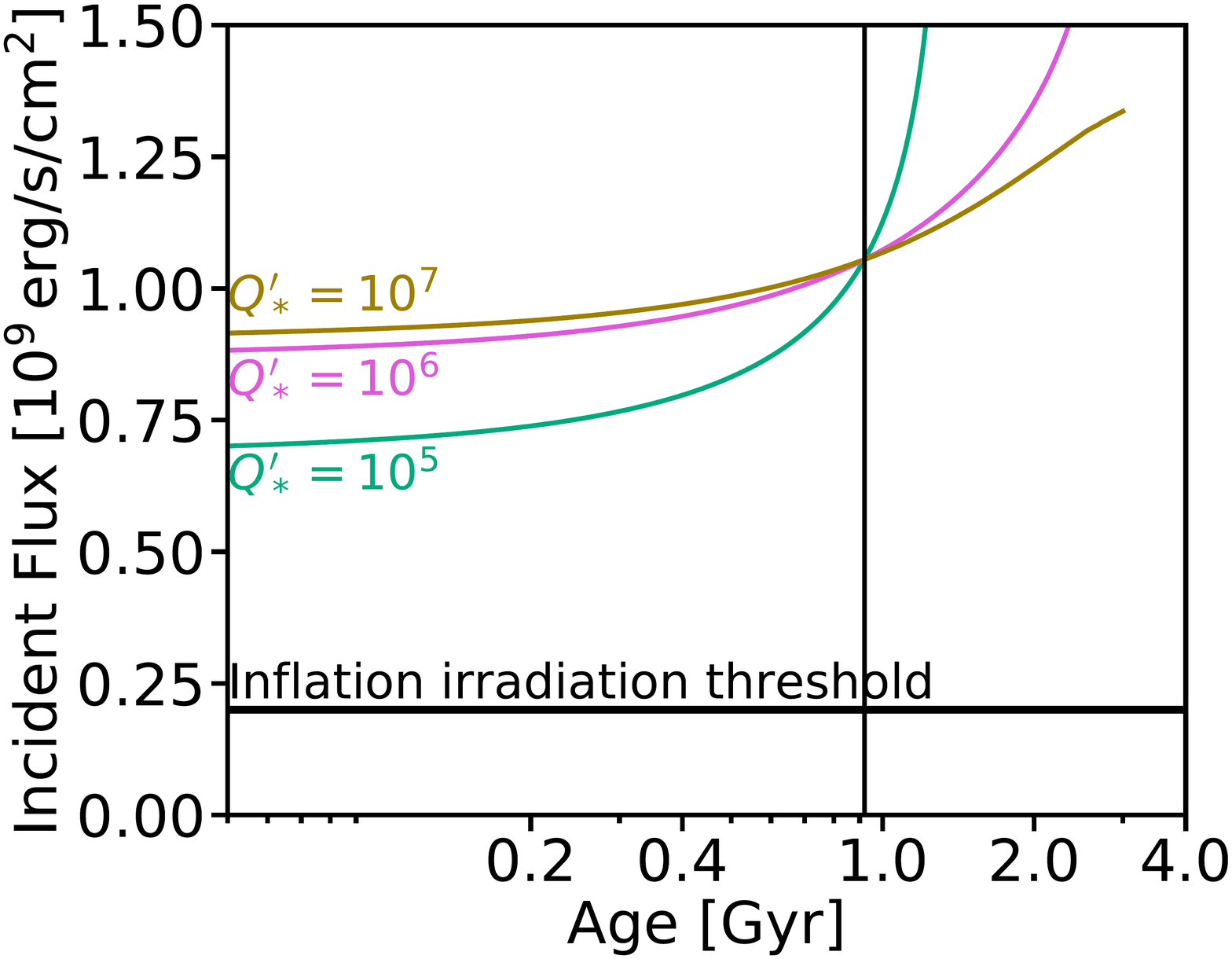}
\caption{Evolution of the semi-major axis (\textit{Top}) and irradiation (\textit{bottom}) for \thisplanet shown for a range of values for $Q^{\star}$. The color of the line indicates the dissipation in the star (green: Q$_{\star}=10^5$, lavender: Q$_{\star}=10^7$, gold: Q$_{\star}=10^8$).
}
\label{fig:evo} 
\end{figure}

We calculated the past and future orbital evolution of the orbit of \thisplanet under the influence of tides, using the POET code \citep{Penev:2014}. We calculated the evolution of the orbital semi-major axis (see Figure \ref{fig:evo}) under the assumptions of a constant tidal phase lag (or constant tidal quality factor), circular orbit, and no perturbations due to further, undetected, objects in the system. Under these assumptions, the tides that the star raises on the planet have no appreciable effect on the orbit, since the angular momentum that can be stored/extracted from the planet is a negligible fraction of the total orbital angular momentum. As a result, the tidal evolution is dominated by the dissipation of tidal perturbations in the star. We accounted for the evolution of the stellar radius, assuming a MIST \citep{Dotter:2016, Choi:2016} stellar evolutionary track appropriate for the best-fit stellar mass and metallicity from our global fit (see \S\ref{sec:GlobalModel}). Finally, we combined the evolution of the orbital semi-major axis with the evolution of the stellar luminosity per the same MIST model to calculate the evolution of the amount of irradiation received by the planet (see Figure \ref{fig:evo}). Because the tidal dissipation in stars is poorly constrained, and likely not well described by a simple constant phase lag model, we considered a broad range of plausible phase lags, parametrized by the commonly used tidal dissipation parameter $Q_{\star}^{\prime}$ (the ratio of the tidal quality factor $Q_{\star}$ and the Love number, k2). 

Regardless of the tidal quality factor, we concluded that the planet has always been subject to a level of irradiation several times larger than the 2$\times$10$^8$ erg s$^{-1}$ cm$^{-2}$ threshold \citet{Demory:2011} suggest is required for the planet to be significantly inflated. Also, again regardless of the amount of dissipation, the planet has undergone at most moderate orbital evolution prior to its current, nearly circular orbit. In contrast, the future fate of the planet is significantly impacted by the amount of tidal dissipation assumed. For tidal quality factor of $Q_{\star}^{\prime}$ = 10$^5$, the planet will be engulfed by its parent star within a few hundred Myrs, while for $Q_{\star}^{\prime}$=10$^7$ or larger the planet survives until the end of the main sequence life of its parent star.

\begin{figure}[ht!]
    \vspace{.0in}
    \centering\includegraphics[width=0.99\columnwidth, trim = 0.0in 0.0in 0.0in 0.0in]{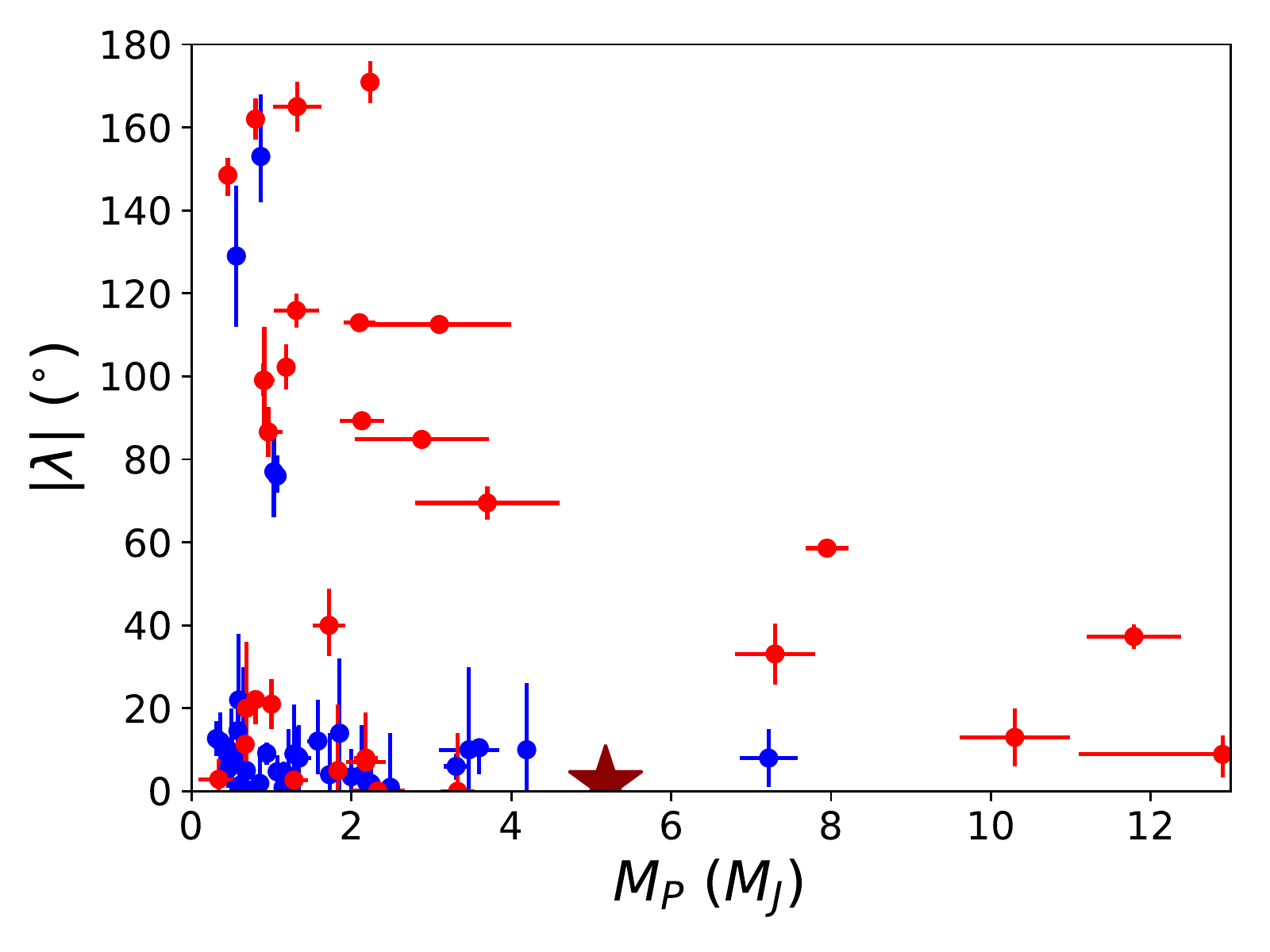}
    \caption{The spin-orbit misalignment of KELT-24b in context with the population of hot Jupiters from the literature. We show the sky-projected spin-orbit misalignments $|\lambda|$ as a function of planetary mass. Red and blue plot points denote planets orbiting stars with effective temperatures less than or greater than the Kraft break at 6250 K, respectively; hot Jupiters orbiting cooler stars typically have well-aligned orbits, whereas those orbiting hotter stars like KELT-24 have a wide range of misalignments \citep{Winn:2010}. We highlight KELT-24b as the large dark red star; the uncertainties are smaller than the plot symbol size. We show only planets with measured masses rather than upper limits, and uncertainties on the spin-orbit misalignments of less than $20^{\circ}$.}
    \label{fig:massvslambda}
\end{figure}

\subsection{\thisstar's Aligned Orbit}

\thisplanet's aligned orbit is interesting in the context of its mass, possible small eccentricity, and the young age of the system. \cite{Hebrard:2010} noted that for massive hot Jupiters, their orbits are typically prograde but with a non-zero misalignment angle, a pattern that still holds true today (see Figure \ref{fig:massvslambda}). \thisplanet is therefore somewhat unusual in that its sky-projected spin-orbit misalignment $\lambda$ is consistent with zero, although the true 3-dimensional spin-orbit misalignment $\psi$ could be larger if the host star is not viewed equator-on. We cannot measure the inclination of the stellar rotation axis $I_{\star}$ using our current data, but a TESS measurement of the rotation period via spot modulation or asteroseismology 
could allow this measurement.

Furthermore, \thisstar's young age and slightly eccentric, aligned orbit place some constraints upon the past history of the system. Some of the high-eccentricity migration mechanisms, such as the Kozai-Lidov mechanism \citep{Anderson:2016} or secular planet-planet interactions \citep{Petrovich:2016} may take hundreds of Myr and typically leave planets in highly eccentric, misaligned orbits. Together with the long tidal damping timescale for the system (longer than the age of the universe), this suggests that \thisplanet likely instead migrated through a faster, less dynamically violent mechanism such as interactions with the protoplanetary disk or in situ formation. 

\subsection{Atmospheric Characterization Prospects}

As mentioned, \thisstar b is one of the few known massive giant planets orbiting a host star bright enough to allow for detailed atmospheric characterization observations. The other comparable planets in this mass range that have been observed with either Spitzer or HST are HAT-P-2 b \citep[Spitzer,][]{lewis:2014}, WASP-14 b \citep[Spitzer,][]{wong:2015}, Kepler-13A b \citep[Spitzer and HST,][]{beatty:2017}, KELT-1b \citep[Spitzer,][]{beatty:2018}, and WASP-103b \citep[Spitzer and HST,][]{kreidberg:2018}. Interestingly, \thisstar b orbits the brightest host in this regime and has the lowest blackbody equilibrium temperature of all these planets: approximately 1450\,K. This places \thisstar b in a different and potentially interesting atmospheric regime. Given the similarities between HAT-P-2b and \thisstar, future atmospheric observation \thisstar b would provide a nice comparison to those already taken for HAT-P-2b.

Observations of massive field brown dwarfs have shown that there is a strong blueward shift in the NIR colors of these objects as they cool from roughly 1400\,K down to approximately 1000\,K. This is known as the ``L-T'' transition, and is generally believed to represent the clouds in the atmospheres of the hotter L-dwarfs slowly dropping below the level of the photosphere in the cooler T-dwarfs. The few observations we have of giant exoplanets in this regime indicate that this transition may occur at cooler temperatures, presumably because the lower surface gravity of the planets is altering the cloud dynamics in their atmospheres, perhaps allowing vertical lofting to maintain the clouds higher for longer \citep{triaud:2015}.

\thisstar b possesses an intermediate surface gravity, three times higher than Jupiter but ten times lower than a brown dwarf, that straddles previous observations. The characterization of the global cloud properties on \thisstar b therefore could allow us to better understand the dynamical processes behind the L-T transition. In particular, a recent analysis of Spitzer phase-curve results by \cite{beatty:2018} has shown that all hot Jupiters appear to posses a nightside cloud deck at a temperature of roughly 1000\,K. The relatively low equilibrium temperature of \thisstar b's atmosphere indicates that even dayside clouds on \thisstar b would be close in composition to the universal nightside clouds on other hot Jupiters. The spectroscopic measurement of \thisstar b's emission might, therefore, be able to determine the specific composition of these clouds. Cloud compositions would in turn provide invaluable insight into the cloud condensation processes, and hence dynamics. 

\section{Conclusion}
\label{sec:conclusion}
We present the discovery of \thisplanet, a massive hot Jupiter in a 5.6-day orbit around a young F-star. The host star, HD 93148, is a bright ($V$=8.3 mag) young F-star (M$_{\star}$ = \starmass\ \msun, R$_{\star}$ = \starradius\ \rsun, age = \starage\ Gyr) with a moderate rotation of $vsinI_*$ = \starvsini\ \kms. The planet is a massive hot Jupiter (M$_P$ = \pmass\ \mj) on a nearly circular ($e$ = \pecc) prograde orbit ($\lambda$ = \plambda\ degrees). \thisstar\ is the brightest host star with a transiting massive Jupiter (M$_P > 4$ \mj), making it well-suited for detailed characterization through eclipse spectroscopy with current facilities like the Hubble Space Telescope and upcoming observatories such as the James Webb Space Telescope. \thisplanet is expected to be observed by NASA's {\it TESS} mission during sectors 20 and 21, from late UT 2019 December to the end of UT 2020 February. This presents a unique opportunity to simultaneously observe the {\it TESS} transits of \thisplanet with Spitzer, HST, and/or ground-based facilities. Additionally, TESS may be able to measure the rotation period of \thisstar, in which case the true (3-D) obliquity can be inferred.

\software{EXOFASTv2 \citep{Eastman:2013, Eastman:2017}, AstroImageJ \citep{Collins:2017}, SPC \citep{Buchhave:2010}}
\facilities{FLWO 1.5m (Tillinghast Reflector Echelle Spectrograph, TRES); Kilodegree Extremely Little Telescope (KELT); MINiature Exoplanet Radial Velocity Array (MINERVA); Las Cumbres Observatory at Tenerife (LCO TFN); University of Louisville Manner Telescope (ULMT, Mt. Lemmon); KeplerCam (FLWO 1.2m); Stacja Obserwacji Tranzyt\'{o}w Egzoplanet w Suwa\l{}kach (SOTES); CROW Observatory; Koyama Astronomical Observatory (KAO)}

\acknowledgements

We thank Laura Kreidberg, Andrew Vanderburg, and James Kirk for their valuable discussions and insight. J.E.R. was supported by the Harvard Future Faculty Leaders Postdoctoral fellowship. Work by G.Z. is provided by NASA through Hubble Fellowship grant HST- HF2-51402.001-A awarded by the Space Telescope Science Institute, which is operated by the Association of Universities for Research in Astronomy, Inc., for NASA, under contract NAS 5-26555. K.G.S. acknowledge support from the Vanderbilt Office of the Provost through the Vanderbilt Initiative in Data-intensive Astrophysics. T.N and A.Y. are also grateful to Mizuki Isogai, Akira Arai, and Hideyo Kawakita for their technical support on observations at Koyama Astronomical Observatory. This work is partly supported by JSPS KAKENHI Grant Numbers JP18H01265 and JP18H05439, and JST PRESTO Grant Number JPMJPR1775. J.L.-B. acknowledges support from FAPESP (grant 2017/23731-1). K.P. acknowledges support from NASA grants 80NSSC18K1009 and NNX17AB94G.

This work makes use of observations from the LCOGT network. This research has made use of SAO/NASA's Astrophysics Data System Bibliographic Services. This research has made use of the SIMBAD database, operated at CDS, Strasbourg, France. This work has made use of data from the European Space Agency (ESA) mission {\it Gaia} (\url{https://www.cosmos.esa.int/gaia}), processed by the {\it Gaia} Data Processing and Analysis Consortium (DPAC, \url{https://www.cosmos.esa.int/web/gaia/dpac/consortium}). Funding for the DPAC has been provided by national institutions, in particular the institutions participating in the {\it Gaia} Multilateral Agreement. This work makes use of observations from the LCO network. This research has made use of the NASA Exoplanet Archive, which is operated by the California Institute of Technology, under contract with the National Aeronautics and Space Administration under the Exoplanet Exploration Program.

MINERVA is a collaboration among the Harvard-Smithsonian Center for Astrophysics, The Pennsylvania State University, the University of Montana, and the University of Southern Queensland. MINERVA is made possible by generous contributions from its collaborating institutions and Mt. Cuba Astronomical Foundation, The David \& Lucile Packard Foundation, National Aeronautics and Space Administration (EPSCOR grant NNX13AM97A), The Australian Research Council (LIEF grant LE140100050), and the National Science Foundation (grants 1516242 and 1608203). Any opinions, findings, and conclusions or recommendations expressed are those of the author and do not necessarily reflect the views of the National Science Foundation. Funding for MINERVA data-analysis software development is provided through a subaward under NASA award MT-13-EPSCoR-0011.

The Center for Exoplanets and Habitable Worlds is supported by the Pennsylvania State University, the Eberly College of Science, and the Pennsylvania Space Grant Consortium.

The ASAS-SN observations are used to help vet exoplanet candidates from KELT. ASAS-SN would like to thank Las Cumbres Observatory and its staff for their continued support. ASAS-SN is funded in part by the Gordon and Betty Moore Foundation through grant GBMF5490 to the Ohio State University, NSF grant AST-1515927, the Mt. Cuba Astronomical Foundation, the Center for Cosmology and AstroParticle Physics (CCAPP) at OSU, the Chinese Academy of Sciences South America Center for Astronomy (CASSACA), and the Villum Fonden (Denmark).

\bibliographystyle{apj}

\bibliography{KELT-24}



\end{document}